\documentstyle{mn}

\begin{document}  

\input epsf
  
\title{THE V R I COLOURS OF HII GALAXIES}
  
\author[Telles \& Terlevich] {Eduardo Telles$^{1,2}$ 
\thanks{present address:
Instituto Astron\^omico e Geof\'{\i}sico - USP,
Caixa Postal 9638,
01065-970 - S\~ao Paulo - BRASIL}
\and Roberto Terlevich$^2$ \\ 
1. Institute of Astronomy, Madingley Road, Cambridge  
CB3 0HA, U.K. \\  2. Royal Greenwich  
Observatory, Madingley Road, Cambridge CB3 0EZ, U.K.\\
etelles@cosmos.iagusp.usp.br, rjt@ast.cam.ac.uk}

\date{accepted on October 8th, 1996.}

\maketitle

\def\cao{\c{c}\~ao~}
\def\CAO{\c{C}\~AO~}
\def\coes{\c{c}\~oes~}
\def\COES{\c{C}\~OES~}
%
%
\def\a4{\hsize 15.cm \vsize 23.cm}
\def\skipline{\vskip 10.1pt}
\def\double{\baselineskip 18pt \lineskip 10pt}     
\def\semidouble{\baselineskip 13pt \lineskip 6pt}
\def\single{\baselineskip 12pt \lineskip 1pt}
\def\supersingle{\baselineskip 10pt \lineskip 1pt}
\def\newpage{\vfill\eject}
%
%
\def\Diaz{\hbox{D\'\i az}}
\def\Vilchez{\hbox{V\'\i lchez}}
\def\neq{\mathrel{\not=}}
\def\twiddles{\hbox{$\sim $}}
\def\varomega{\varpi}
%
%
\def\Angeles{\hbox{Angeles I. D\'\i az}}
\def\Pepe{\hbox{Jos\'e M. V\'\i lchez}}
\def\Elena{\hbox{Elena Terlevich}}
\def\Bernard{\hbox{Bernard E.J. Pagel}}
\def\Mike{\hbox{Michael G. Edmunds}}
%
%
\def\ang{\thinspace\hbox{\AA}}
\def\km{\thinspace\hbox{km}}
\def\Mpc{\thinspace\hbox{Mpc}}
\def\kpc{\thinspace\hbox{kpc}}
\def\kmsec{\thinspace\hbox{$\hbox{km}\thinspace\hbox{s}^{-1}$}}
\def\kmsecmeg{\thinspace\kmsec\Mpc$^{-1}$}
\def\ergsec{\thinspace\hbox{$\hbox{erg}\thinspace\hbox{s}^{-1}$}}
\def\ergsqcmsec{\thinspace\hbox{erg}\sqcm\sec$^{-1}$}
\def\sqcm{\thinspace\hbox{$\hbox{cm}^{2}$}}
\def\cm3{\thinspace\hbox{$\hbox{cm}^{3}$}}
\def\cm2{\thinspace\hbox{$\hbox{cm}^{2}$}}
\def\persqcm{\thinspace\hbox{$\hbox{cm}^{-2}$}}
\def\percucm{\thinspace\hbox{$\hbox{cm}^{-3}$}}
\def\kev{\thinspace\hbox{keV}}
\def\sec{\thinspace\hbox{s}}
\def\ha{\hbox{$\hbox{H}_\alpha$}}
\def\hb{\hbox{$\hbox{H}_\beta$}}
\def\hg{\hbox{$\hbox{H}_\gamma$}}
\def\hd{\hbox{$\hbox{H}_\delta$}}
\def\lya{\hbox{$\hbox{Ly}_\alpha$}}
\def\kelvin{\thinspace\hbox{K}}
\def\dyncm2{\thinspace\hbox{$\hbox{dyn}\thinspace\hbox{cm}^{-2}$}}
\def\deg{\hbox{$^\circ$}}
\def\rstar{\thinspace\hbox{$\hbox{R}_*$}}
\def\vstar{\thinspace\hbox{$\hbox{V}_*$}}
\def\Zsun{\thinspace\hbox{$\hbox{Z}_{\odot}$}}
\def\msun{\thinspace\hbox{$\hbox{M}_{\odot}$}}
\def\rsun{\thinspace\hbox{$\hbox{R}_{\odot}$}}
\def\lsun{\thinspace\hbox{$\hbox{L}_{\odot}$}}
\def\gpar{\hbox{$g_{\parallel}$}}
\def\Ca{$\lambda\lambda$\thinspace\hbox{8498,8542,8662}~\AA}
\def\C2{$\lambda\lambda$\thinspace\hbox{8542,8662}~\AA}
\def\Mg{$\lambda$\thinspace\hbox{8807}~\AA}
\def\Fe{\hbox{$[Fe/H]$}}
\def\lg{\hbox{$log\thinspace\hbox{g}$}}
\def\lT{\hbox{$log\thinspace\hbox{T}_{eff}$}}
\def\Teff{\hbox{$T_{eff}$}}
%
%
\def\UAM {\ninesl {Depto. de F\'\i sica Te\'orica, C-XI, 
Univ. Aut\'onoma de Madrid, 28049 Madrid, Spain.}}
\def\RGO {\ninesl{Royal Greenwich Observatory, Madingley Rd.,
Cambridge, CB3 0EZ, U.K.}}
\def\IAC {\ninesl{Instituto de Astrof\'\i sica de Canarias, 
38200 La Laguna, Tenerife, Spain}}
\def\Cardiff {\ninesl{Department of Physics, University of Wales 
College of Cardiff, PO Box 913, Cardiff, CF1 3TH, U.K.}}
\def\Nordita {\ninesl{NORDITA, Blegdamsvej 17, DK-2100 K\obenhavn \O,
Denmark}}
%
\def\reference#1#2#3#4#5{\tenp\par\noindent\hangindent 3em
              #1, #2. {\tenpsl #3\/}, {\tenpb #4,}
\thinspace\hbox{#5}}
\def\refnum#1#2#3#4{\tenp\noindent #1, #2. {\sl #3\/}, {\bf #4}}
\def\refpress#1#2#3{\tenp\par\noindent\hangindent 3em
             #1, #2. {\tenpsl #3\/}, in press}
\def\refsub#1#2#3{\tenp\par\noindent\hangindent 3em
           #1, #2. {\tenpsl #3\/}, submitted}
\def\refprep#1#2#3{\tenp\par\noindent\hangindent 3em
           #1, #2. {\tenpsl #3\/}, in preparation}
\def\refacc#1#2#3{\tenp\par\noindent\hangindent 3em
           #1, #2. {\tenpsl #3\/}, accepted}
\def\refbook#1#2#3{\tenp\par\noindent\hangindent 3em
            #1, #2. {\tenpsl #3\/}}

\def\etal   {et\nobreak\ al.\ }
\def\etalp   {{\tenib et\nobreak\ al.\ }}
\def\etalr   {{\tenpsl et\nobreak\ al.\ }}
\def\aanda  {Astr.\ Astro\-phys.\nobreak\ }
\def\aandas {Astr.\ Astro\-phys.\nobreak\ Suppl.\nobreak\ }
\def\aandal {Astr.\ Astro\-phys.\nobreak\ Lett.\nobreak\ }
\def\aj     {Astron.\nobreak\ J.\nobreak\ }
\def\annrev {Ann.\ Rev.\ Astr.\ Astro\-phys.\nobreak\ }
\def\acta   {Acta Astron.\nobreak\ }
\def\apj    {Astro\-phys.\nobreak\ J.\nobreak\ }
\def\apjs   {Astro\-phys.\nobreak\ J.\ Suppl.\nobreak\ }
\def\apjl   {Astro\-phys.\nobreak\ J.\ Lett.\nobreak\ }
\def\apspsc {Astro\-phys.\nobreak\ Sp.\nobreak\ Sc.\nobreak\ }
\def\apjsupp{\apjs}
\def\aplett {Astro\-phys.\nobreak\ Lett.}
\def\commap {Comments\ Astrophys.\nobreak\ }
\def\ibvs   {Inf.\ Bull.\ var.\ Stars}
\def\mn     {Mon.\ Not.\ R.\ astr.\nobreak\ Soc.\nobreak\ }
\def\pasp   {Publ.\ astr.\ Soc.\ Pacif.\nobreak\ }
\def\pasj   {Publ.\ astr.\ Soc.\ Japan\nobreak\ }
\def\sovast {Soviet astr.}
\def\rmx    {Rev.\ Mex.\ Astr.\ Astrofis.\nobreak\ }
\def\jump   {\par\vskip 0.2cm\noindent\hangindent 3em}
%
%
\newcount\notenumber
\def\clearnotenumber{\notenumber=0}
\def\note{\advance\notenumber by 1
  \footnote{$^{\the\notenumber}$}}
\newcount\natrefreg
\def\clearnatref{\natrefreg=0}
\def\natref{\advance\natrefreg by 1 $^{\the\natrefreg}$}
\newcount\secreg
\def\clearsecreg{\secreg=0}
\newcount\subsecreg
\def\clearsubsecreg{\subsecreg=0}
\def\newsection#1{\skipline\skipline\par\smallbreak\noindent\advance\secreg by1
 \the\secreg . {\bf #1}\nobreak\par\clearsubsecreg}
\newcount\tablereg
\def\cleartable{\tablereg=0}
\def\nextable{\advance\tablereg by1\table}
\newcount\figreg
\def\clearfig{\figreg=0}
\def\fig{Fig.~\the\figreg}
\def\nextfig{\advance\figreg by1\fig}
%
%
\def\rsa{{\tenpsl Revised Shap\-ley-Ames Catalog of Bright Gal\-axies\/} Sandage
 \&\ Tammann (1981)}
\def\RC2{{\tenpsl Second Reference Catalogue of Bright Gal\-axies\/},
 de~Vaucouleurs, de~Vaucouleurs and Corwin (1976)}
\def\rc2{\RC2}
\def\Sin{\hbox{Sin}}
\def\Cos{\hbox{Cos}}
%
%
\def\deriv#1#2{\hbox{${{\displaystyle\hbox{d}#1}\over
{\displaystyle\hbox{d}#2}}$}}
\def\sderiv#1#2{\hbox{${{\displaystyle\hbox{d}^2#1}\over
{\displaystyle\hbox{d}#2^2}}$}}
\def\pderiv#1#2{\hbox{${{\displaystyle\partial#1}\over
{\displaystyle\partial#2}}$}}
\def\spderiv#1#2{\hbox{${{\displaystyle\partial^2#1}\over
{\displaystyle\partial#2^2}}$}}
\def\x10#1{\hbox{$\times\hbox{10}^{#1}$}}
\def\eex#1{\hbox{$\hbox{10}^{#1}$}}
\def\movements#1{\halign{\quad\it##\hfil&&\qquad\it##\hfil\cr#1\crcr}}
%
%
\def\integ{\int\limits}
%
%
\font\title=cmbx10 scaled 1440
\font\subt=cmbx10 scaled 1200
\font\tenp=cmr10
\font\poster=cmr10 scaled 1440
\font\tenpoint=cmr10 scaled 1200
\font\nine=cmr9
\font\ninei=cmti9
\font\nineb=cmbx9
\font\tenpib=cmbxti10
\font\tenib=cmbxti10 scaled 1200
\font\ninesl=cmsl9
\font\mit=cmmi10 scaled 1200
\font\tenpsl=cmsl10
\font\tenpb=cmbx10  
  
\begin{abstract}  

We present a high spatial resolution CCD surface photometry study in the
optical V, R and I broadband filters of a sample of 15 HII galaxies.
Narrow band imaging has allowed the separation of the emission line
region from the extended parts of the galaxy.  The latter are assumed
to represent the underlying galaxy in HII galaxies, thus, the colours
of the underlying galaxy are measured.  The colours of the underlying
stellar continuum within the starburst are also derived by subtracting
the contribution of the emission lines falling in the broad band
filters.  The distribution of colours of the underlying galaxy in HII
galaxies is similar to the colours of other late type low surface
brightness galaxies which suggests a close kinship of these with the
quiescent phases of HII galaxies.  However, comparison with recent
evolutionary population synthesis models shows that the observational
errors and the uncertainties in the models are still too large to put
strict constraints on their past star formation history.

Our analysis of the morphology and structural properties, from contour
maps and luminosity profiles, of this sample of 15 HII galaxies agree
to what has been found in Telles (1995) and Telles, Melnick \&
Terlevich (1996), namely that, HII galaxies comprise two broad
classes segregated by their luminosity; Type I HII galaxies are
luminous and have disturbed and irregular outer shapes while Type II
HII galaxies are less luminous and have regular shapes.  The outer
parts of their profiles are well represented by an exponential as in
other types of known dwarf galaxy.
\end{abstract}

\section{Introduction}

The question of whether HII galaxies are primordial galaxies
experiencing their very first burst of star formation or if an older
stellar population from an earlier event of star formation is present
has not, since it was first posed by Sargent \& Searle (1970), been
answered. Multi-colour surface photometry can provide with the
answers.  To infer ages and mix of stellar populations of HII galaxies
through broad band observations one can compare the observed colours
with those derived from evolutionary synthesis models.  When dealing
with systems with intense gaseous emission though, line emission can
be a significant contributor to the total fluxes of the starburst
within the corresponding broad band filters (Huchra 1977; Salzer,
MacAlpine \& Boroson 1989).  In this case, two approaches may be
considered.  On one hand, one may compare the integrated colours of
the starburst with synthesis models which take into account the
gaseous line emission (Huchra 1977) or use stellar evolutionary models
by comparing with the starburst stellar continuum alone.  The latter
can be derived by accounting for the contribution of the emission
lines in the broad band filters of interest.  We have chosen the
second approach relying on the spectrophotometric information from
{\em The Spectrophotometric Catalogue of HII Galaxies} (Terlevich
\etal 1991, hereafter SCHG) and some assumption about the gas emission
line surface distribution and the availability of state of the art
stellar evolutionary population synthesis models.

In this paper, we compare the results of high spatial resolution CCD
broadband optical photometry with the observed colours of samples of
dwarf galaxies and with recent models of evolutionary synthesis from
the literature. In the case of the starburst region in HII galaxies
the combined effect of low mass stars, if present from previous
episodes of star formation, on the optical colours is expected to be
negligible relative to the dominant contribution of the high mass
stars responsible for the observed ionized spectrum.  A positive
direct detection of an older stellar population in the burst region is
a very difficult task.  We take advantage of the high spatial
resolution observations that allow us to estimate the colours beyond
the line emitting regions which we will call extensions.  The
extensions are virtually free from line emission, therefore the light
in these regions is assumed to be produced by a population of stars
other than that embedded in the starburst.  Thus, we put constraints on
the ages of the underlying galaxy by comparing the colours of these
extensions with the models and with control samples of real galaxies.

Through the means of surface photometry (e.g. morphology, structure,
luminosity profiles and colours) we re-address the question of what HII
galaxies resemble in their quiescent period and their possible kinship
with other known types of dwarf galaxies.

We present the sample selection in Section~\ref{stellar:sample},
observations and data reduction in Section~\ref{stellar:reduction}.
Section~\ref{stellar:results} shows all the results of the surface
photometry. In \S\,\ref{stellar:discuss} we discuss various topics
raised by our data.  Finally, in Section~\ref{stellar:conclusions}, we
present our conclusions.

\section{The Sample}\label{stellar:sample}

\begin{table*}
\centering
\caption[Positions and basic spectroscopic data for the present galaxy
sample.]{Positions and basic spectroscopic data for the galaxy sample.
The data are from SCHG and from Melnick, Terlevich \& Moles (1988).
Oxygen abundance determinations are from: $^\dagger$ Melnick, Terlevich \&
Moles (1988); $^\ddagger$ Telles (1995) using procedure as
in Pagel \etal (1992); $^\amalg$ R$_{23}$ empirical calibration from
McGaugh (1991); $^\ast$ Gonz\'alez-Delgado \etal (1995).}
\label{stellar:positions}
\begin{tabular}{lrrcccccc} \hline
{\em name} & R.A.(1950) & $\delta$ (1950) &  z  &  $\sigma$ & $
{\cal F}(H\beta)$ &  C$(H\beta)$ & W$(H\beta)$ &  O/H            \\ \hline
UM 238        & 00 22 06.0 & 01 27 36  & 0.014 &      &  0.08 & 0.00  & 149 & 8.14$^\ddagger$ \\
UM 133        & 01 42 08.0 & 04 38 48  & 0.009 & 17.2 &  0.32 & 0.43 & 65  & 7.69$^\ddagger$ \\
UM 408        & 02 08 48.0 & 02 06 36  & 0.012 &      &  0.07 & 0.62 & 52  & 7.66$^\ddagger$   \\
II Zw 40      & 05 53 00.6 & 03 24 00  & 0.003 & 35.2 &  1.90 & 1.00 & 170 & 8.13$^\dagger$   \\
C0840+1044 & 08 39 53.2 & 10 44 02  & 0.012 & 34.0 &  0.10 & 0.38 & 55  &    \\
C0840+1201 & 08 39 36.3 & 12 00 49  & 0.030 & 36.5 &  0.48 & 0.52 & 105 & 7.88$^\dagger$ \\
C08-28A    & 08 42 45.4 & 16 16 46  & 0.054 & 49.1 &  0.28 & 0.77 & 35  & 8.40$^\amalg$  \\
Mrk 36        & 11 02 01.2 & 29 24 00  & 0.002 & 16.0 &  1.40 & 0.70 & 70  & 7.86$^\dagger$   \\
UM 448        & 11 39 38.3 & 00 36 38  & 0.018 & 40.8 &  2.20 & 0.48 & 45  & 8.03$^\ddagger$ \\
UM 455        & 11 47 50.0 & -00 15 01 & 0.012 & 20.6 &  0.09 & 0.26 & 55  & 7.84$^\dagger$    \\
UM 461A        & 11 48 59.4 & -02 05 41 & 0.005 & 14.5 &  0.70 & 0.40 & 155 & 7.74$^\dagger$    \\
UM 483        & 12 09 41.0 & 00 21 00  & 0.007 &      &  0.14 & 0.11 & 26  & 8.24$^\ddagger$ \\
C1212+1148 & 12 10 56.5 & 11 57 36  & 0.023 & 34.2 &  0.13 & 0.63 & 60  & 7.98$^\amalg$ \\
C1409+1200 & 14 09 14.3 & 11 59 33  & 0.056 & 52.3 &  0.19 & 0.44 & 130 & 8.18$^\dagger$ \\
UM 167        & 23 33 40.0 & 01 52 30  & 0.009 &      &  2.51 & 0.26 & 28  & 8.50$^\ast$ \\
\hline
\end{tabular}
\end{table*}

The sample we selected for the surface photometry study is a
sub-sample of SCHG for which emission line width measurements are, in
most cases, available in the literature (Melnick, Terlevich \& Moles
1988).  A recent review of the spectroscopic properties of the HII
galaxies in SCHG is given by Telles (1995, and references therein).
We have compiled in Table~\ref{stellar:positions} some of the basic
spectroscopic data of the present galaxy sample.  Column 1 gives the
name of the objects used in this study, while columns 2 and 3 give
their 1950 position.  Column 4 gives the redshift (z) and 5 gives the
emission line width ($\sigma$) in \kmsec, as given by Melnick,
Terlevich \& Moles (1988).  Column 6 gives the Flux in H$\beta$
[${\cal F}(H\beta)$] in units of 10$^{-13}$ \ergsqcmsec, 7 gives the
observed reddening coefficient C$(H\beta)$ obtained from the Balmer
decrement and column 8 the equivalent width of H$\beta$ [W$(H\beta)$]
in \AA.  The last column gives the oxygen abundance in the units of
12+log(O/H).  Whenever we did not find in the literature, we
calculated the oxygen abundances using the prescriptions given in
Pagel \etal (1992) in case the [OIII] $\lambda4363$ line was detected.
Otherwise, we have used the empirical abundance estimator devised by
Pagel \etal (1979) with the calibration of McGaugh (1991).

\section{Observations and Data Reduction}\label{stellar:reduction}

\begin{table*}
\centering
\caption{Journal of Observations.}
\label{stellar:journal}
\begin{tabular}{lccccccc} \hline
{\em name} & instrument & date & CCD & pixel & 
\multicolumn{3}{c}{Exposure time} \\
 & & & & size ($\,''$)& V & R & I \\ \hline
UM 238        & NOT2.5m & 13JAN91 & TK512-011 & 0.20 & 1000 & 1000 & 1200 \\
UM 448        & NOT2.5m & 13JAN91 & TK512-011 & 0.20 & 1000 & 1000 & 1000 \\
II Zw 40      & NOT2.5m & 13JAN91 & TK512-011 & 0.20 & 1000 & 1000 & 500+500 \\
C08-28A    & NOT2.5m & 14JAN91 & TK512-011 & 0.20 & 1000 & 1000 & 1000 \\
Mrk 36        & NOT2.5m & 14JAN91 & TK512-011 & 0.20 & 1000 & 1000 & 1000 \\
UM 455        & NOT2.5m & 14JAN91 & TK512-011 & 0.20 & 1000 & 1000 &  \\
UM 133        & NOT2.5m & 15JAN91 & TK512-011 & 0.20 & 1000 & 1000 & 1000 \\
UM 408        & NOT2.5m & 15JAN91 & TK512-011 & 0.20 & 1000 & 1000 & 1000 \\
C0840+1044 & NOT2.5m & 15JAN91 & TK512-011 & 0.20 & 1000 & 1000 & 1000 \\
UM 461        & NOT2.5m & 15JAN91 & TK512-011 & 0.20 & 1000 & 1000 & 1000 \\
C1212+1148 & NOT2.5m & 15JAN91 & TK512-011 & 0.20 & 500 & 500 & 500 \\
C0840+1201 & JKT1.0m & 11MAR92 & GEC3 & 0.33 & 1800 & 1800 & 1200+1200 \\
UM 483        & JKT1.0m & 11MAR92 & GEC3 & 0.33 & 1800 & 1800 & 1800 \\
C1409+1200 & JKT1.0m & 11MAR92 & GEC3 & 0.33 & 1800 & 1800 & 1800+1800 \\
UM 167        & JKT1.0m & 25OCT92 & EEV7 & 0.31 & 500+300+200 & 200 & 1800 \\
\hline
\end{tabular}
\end{table*}

The observations were made using the 2.5-m Nordic Optical Telescope
(NOT) and the 1.0-m the Jacobus Kapteyn Telescope (JKT) at the
Observatorio del Roque de Los Muchachos, La Palma, Canary Islands.
The journal of all broadband observations is given in
Table~\ref{stellar:journal}.  
The JKT observations were obtained as part of the GEFE\footnote{GEFE,
Grupo de Estudios de Formaci\'on Estelar, is an international
collaboration of astronomers from Spain, the UK, France, Germany,
Denmark and Italy, formed to take advantage of the international time
granted by the Comit\'e Cient\'{\i}fico Internacional at the
Observatories in the Canary Islands. The aim of this project is the
study of star formation processes in young extragalactic stellar
systems.} project.  The nights were photometric and had typically
sub-arcsecond seeing conditions.  For the NOT observations the
combined optical system at the Cassegrain focal station gives a focal
ratio of f/11.0.  We have used a Tektronix TK512-011 \,``thick\,''
frontside-illuminated 512x512 CCD chip giving a total field of about
$1.7 {\tt '} \times 1.7 {\tt '}$.  Small correction for non-linearity
of this CCD chip had to be applied which should not introduce any
significant error to the photometry.  We used the linearity curve
determined by Hans Kjeldsen (private communication) to perform this
correction. For the JKT observations in March 91 we used the GEC3
400x590 CCD chip which gives a total field of $2.2 {\tt '} \times 3.2
{\tt '}$ at the Cassegrain focal station.  In October, we used the
EEV7 1280x1180 CCD chip which gives a total field of $6.5 {\tt '}
\times 6.0 {\tt '}$.  The small pixel sampling of the observations
(0.2${\tt ''}$ for the NOT and 0.3${\tt ''}$ for the JKT) combined
with the excellent seeing allowed us to obtain images of high spatial
resolution.  We have used broad-band Cousins V, R and I filters and
narrow band filters centered on the redshifted [OIII] (NOT) or
H$\alpha$ (JKT) emission lines.  The latter were used to discriminate
the regions of line emission in the galaxies.  The correction for the
contribution of the emission line to the total fluxes in the broadband
observations is described in appendix A.  For this purpose we needed a
precise knowledge of the combined filter and CCD transmission
responses.  These were obtained from the description of the filters
given in NOTNEWS (1989) and in the Isaac Newton Group manuals.

Basic data reduction consisted of bias subtraction and flat field
division, cosmic rays elimination, using the Image Reduction and
Analysis Facility (IRAF) under Unix at the computing facilities of the
Starlink node at Cambridge.  Sky subtraction was performed by
determining the mean value of several measurements of different areas
surrounding the object.  Various tests were performed to estimate the
accuracy of the sky subtraction which we believe to be better than
1\%.  Flux calibration was done through the observation of photometric
standard stars from Landolt (1983) throughout the nights. Stars were
masked out from the images.  The star-free images were then used to
obtain the integrated magnitudes for each object from the asymptotic
values of the circular aperture curve of growth.

We have determined the region where the gaseous line emission is
dominant.  This region is representative of the ionized nebula in HII
galaxies.  This determination of the burst region has been made
through the use of the narrow band images centered on the redshifted
[OIII] lines for the galaxies in the NOT sample and on the redshifted
H$\alpha$ line for the JKT sample.  In both cases, we discriminate the
burst region of the galaxy as being the isophotal level with signal to
noise ratio (S/N) greater than 2.  We then aligned the [OIII] frames
with the broad band frames and separated the broad band burst region
alone for each object.  Thus, we were able to estimate the fraction of
the total integrated fluxes within burst regions and also the fluxes
in the extensions, tails or fuzz whose colours are unaffected by line
emission.

\section{Results}\label{stellar:results}

\subsection{Integrated photometry}\label{stellar:integrated}

\begin{table*}
\centering
\caption[Observed total Cousins magnitudes and internal
accuracy.]{Observed total Cousins magnitudes and internal accuracy (no
extinction correction).  The last three columns show the total
absolute magnitudes corrected for galactic extinction.}
\label{tab: photo}
\begin{tabular}{lcccccc} \hline
{\em object} & V & R & I & M$_{\rm V}$ & M$_{\rm R}$ & M$_{\rm I}$ \\ \hline
UM238      & 16.52$\pm$0.02 & 16.17$\pm$0.02 & 15.89$\pm$0.02 & -18.10 & -18.45 & -18.73 \\
UM133      & 15.41$\pm$0.03 & 15.07$\pm$0.03 & 14.63$\pm$0.03 & -18.25 & -18.60 & -19.03 \\
UM408      & 17.38$\pm$0.03 & 17.12$\pm$0.03 & 16.73$\pm$0.03 & -16.91 & -17.16 & -17.56 \\
IIZw40     & 14.59$\pm$0.02 & 13.89$\pm$0.02 & 13.66$\pm$0.02 & -18.42 & -18.67 & -18.44 \\
C0840+1040 & 18.00$\pm$0.03 & 17.65$\pm$0.03 & 17.33$\pm$0.03 & -16.37 & -16.71 & -17.00 \\
C0840+1201 & 16.51$\pm$0.05 & 16.21$\pm$0.04 & 16.20$\pm$0.03 & -19.86 & -20.13 & -20.12 \\
C08-28A  & 15.33$\pm$0.04 & 14.92$\pm$0.02 & 14.60$\pm$0.02 & -22.22 & -22.64 & -22.95 \\
Mark36     & 15.25$\pm$0.04 & 14.90$\pm$0.02 & 15.04$\pm$0.02 & -15.15 & -15.49 & -15.36 \\
UM448      & 13.88$\pm$0.02 & 13.36$\pm$0.02 & 13.16$\pm$0.02 & -21.29 & -21.81 & -22.01 \\
UM455      & 16.74$\pm$0.04 & 16.47$\pm$0.02 & 16.23$\pm$0.20 & -17.54 & -17.82 & -18.06 \\
UM461      & 16.10$\pm$0.03 & 15.89$\pm$0.03 & 15.66$\pm$0.03 & -16.29 & -16.50 & -16.73 \\
UM483      & 15.76$\pm$0.05 & 15.46$\pm$0.04 & 15.20$\pm$0.03 & -17.35 & -17.66 & -17.92 \\
C1212+1148 & 17.40$\pm$0.03 & 17.16$\pm$0.03 & 16.69$\pm$0.03 & -18.30 & -18.54 & -19.01 \\
C1409+1200 & 17.45$\pm$0.05 & 17.37$\pm$0.04 & 17.31$\pm$0.03 & -20.18 & -20.26 & -20.32 \\
UM167      & 12.09$\pm$0.04 & 11.74$\pm$0.04 & 11.50$\pm$0.03 & -21.70 & -22.01 & -22.22 \\
\hline
\end{tabular}
\end{table*}

Results of the photometric measurements and {\em internal} accuracy of
photometric calibration for the observations are given in
Table~\ref{tab: photo}.  No correction for internal or external
extinction has been applied for the \underline{observed} magnitudes.
Column 1 gives the name of the galaxy. Columns 2, 3 and 4 give the
integrated apparent V R I magnitudes from the asymptotic value of the
circular aperture curve of growth with their corresponding internal
errors.  Columns 5, 6 and 7 give the derived total absolute magnitudes
after correction for galactic extinction and using a long distance
scale\footnote{H$_0 = 50$ \kmsecmeg~is used throughout this paper.}.
Table~\ref{tab: compare} shows the comparison of our photometry with
values quoted by different authors in the literature for some galaxies
in common.  The agreement is better than 0.07 mag.

\begin{table}
\centering
\caption[Comparison of our photometry with other authors.]{Comparison
of the photometry with other authors. Salzer, MacAlpine \& Boroson
(1989) $^\dagger$, RC3 $^\ddagger$, Mazzarela \& Boroson (1993)
$^\amalg$, Huchra (1977)$^\ast$.}
\label{tab: compare}
\begin{tabular}{lccc} \hline
 {\em object}   &  broad &    Our&  Other \\
                & bandpass & photometry &  authors \\ \hline
II Zw 40 & V & 14.59$\pm$0.02 &  14.66$\pm$0.13$^\ddagger$ \\
UM 448   & V & 13.88$\pm$0.02 &  13.90$\pm$0.05$^\dagger$  \\
         & R & 13.36$\pm$0.02 &  13.38$\pm$0.02$^\amalg$ \\
Mark 36  & V & 15.25$\pm$0.04 &  15.24$\pm$0.05$^\ast$  \\
         & R & 14.90$\pm$0.04 &  14.90$\pm$0.05$^\ast$  \\
UM 455   & V & 16.74$\pm$0.04 &  16.73$\pm$0.13$^\dagger$ \\
UM 408   & V & 17.38$\pm$0.03 &  17.20$\pm$0.06$^\dagger$ \\
UM 461   & V & 16.10$\pm$0.03 &  15.93$\pm$0.06$^\dagger$ \\
UM 483   & V & 15.76$\pm$0.05 &  15.91$\pm$0.30$^\dagger$ \\
\hline
\end{tabular}
\end{table}

For the analysis that follows the apparent magnitudes were then
corrected for Galactic absorption, where appropriate, by using the
maps of Burstein \& Heiles (1982).  IIZw40, a low galactic latitude
object, has the largest correction E(B-V)=0.56; C0840+1044 and
C0840+1200 corrections were estimated to be E(B-V)=0.03; for UM 167
E(B-V)=0.04. The total extinction in different filters were estimated
using A$_{\rm V}$=3.10 E(B-V), A$_{\rm R}$=2.30 E(B-V), A$_{\rm
I}$=1.48 E(B-V) (Varela 1992, and references therein).  No attempt has
been made to correct for internal absorption, $K$-correction or
inclination.

\begin{table}
\centering
\caption[Half-light sizes and mean effective surface
brightness.]{Half-light sizes and mean effective surface brightness.
These values are corrected for galactic extinction.}
\label{tab: effective}
\begin{tabular}{lcccccc} \hline
 {\em object} &\multicolumn{3}{c}{R$_{eff}(\,'')$}
              &\multicolumn{3}{c}{$<\mu>_{eff}$}\\ \hline
  &  V & R & I &  V & R & I  \\ \hline
UM238       &  6.47 &  7.07 &  7.91 & 22.57 & 22.41 & 22.38 \\
UM133       & 17.10 & 17.30 & 17.80 & 23.57 & 23.25 & 22.88 \\
UM408       &  2.07 &  2.14 &  2.44 & 20.95 & 20.77 & 20.66  \\
IIZw40      &  8.20 &  7.30 & 14.60 & 19.42 & 18.92 & 20.65  \\
C0840+1044  &  2.35 &  2.62 &  3.24 & 21.76 & 21.67 & 21.83  \\
C0840+1201  &  2.60 &  2.85 &  3.19 & 20.49 & 20.41 & 20.67  \\
C08-28   &  3.93 &  4.05 &  4.74 & 20.30 & 19.95 & 19.98   \\
Mrk36       &  4.50 &  4.97 &  5.49 & 20.51 & 20.38 & 20.73   \\
UM448       &  4.08 &  4.04 &  3.68 & 18.93 & 18.38 & 17.99  \\
UM455       &  3.14 &  3.18 &       & 21.22 & 20.97 &        \\
UM461       &  3.95 &  4.20 &  6.00 & 21.08 & 21.00 & 21.55 \\
UM483       &  3.50 &  3.58 &  3.52 & 20.48 & 20.23 & 19.93  \\
C1212+1148  &  1.13 &  1.18 &  1.59 & 19.66 & 19.51 & 19.70  \\
C1409+1200  &  1.26 &  1.41 &  1.60 & 19.94 & 20.11 & 20.33  \\
UM167       & 11.20 &  9.47 &  8.20 & 19.20 & 18.53 & 18.00  \\ \hline
\end{tabular}
\end{table}

Table~\ref{tab: effective} shows the {\em dereddened} effective values
measured directly from the curves of growth at half-light ($m_{eff} =
m_{asymp} + 0.7526$) for the three different filters.  The effective
radius $r_{eff}$ is radius which contains half of the total light,
read from the curve of growth at $m_{eff}$.  $<\mu>_{eff}$ is the mean
effective surface brightness which is the mean surface brightness
within the effective radius $r_{eff}$.

\subsection{Contour maps}

\begin{figure*}
\protect\centerline{
\epsfxsize=5.2in\epsffile[ 1 1 710 570]{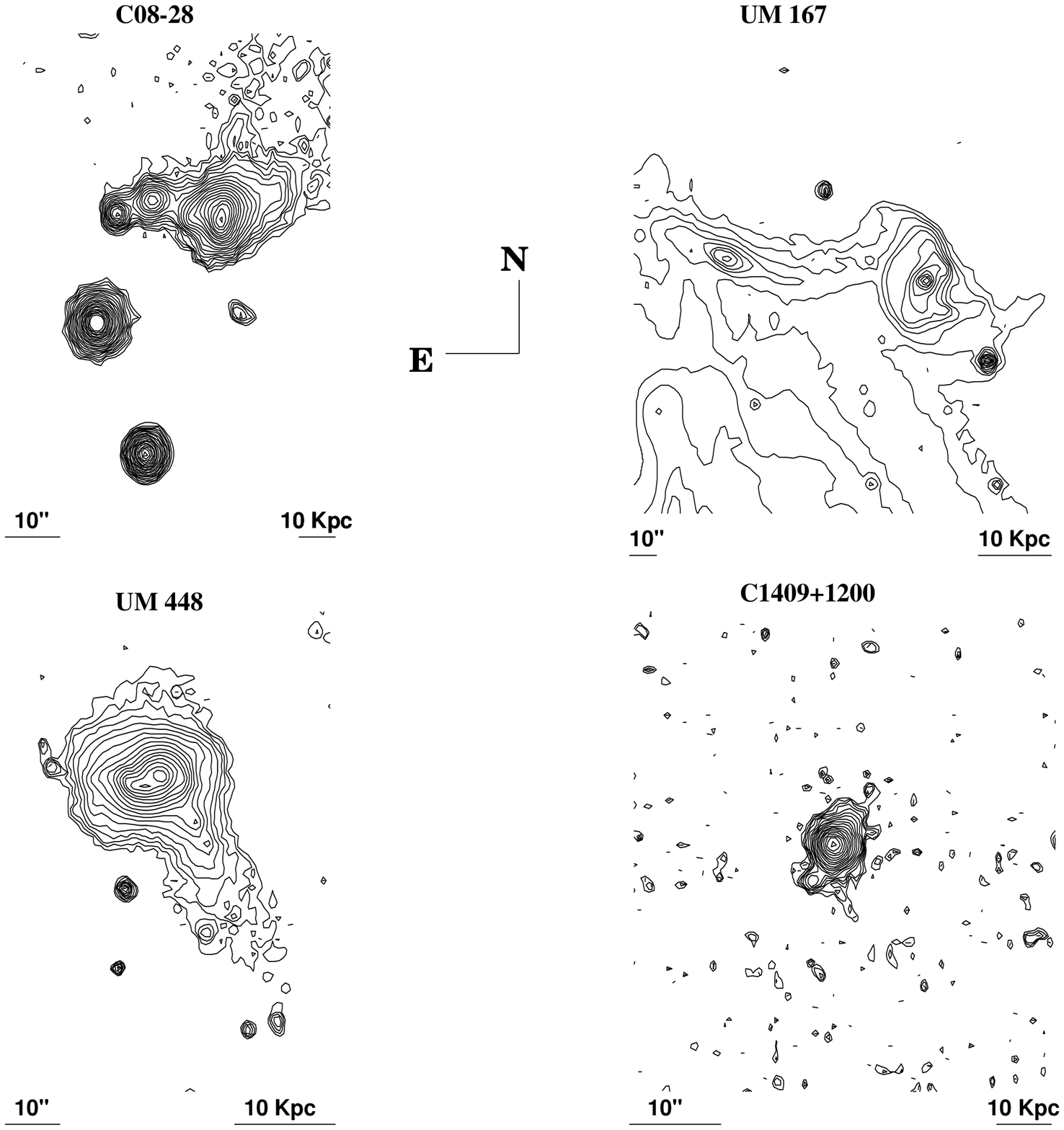}
}
\protect\centerline{
\epsfxsize=5.2in\epsffile[ 1 1 710 570]{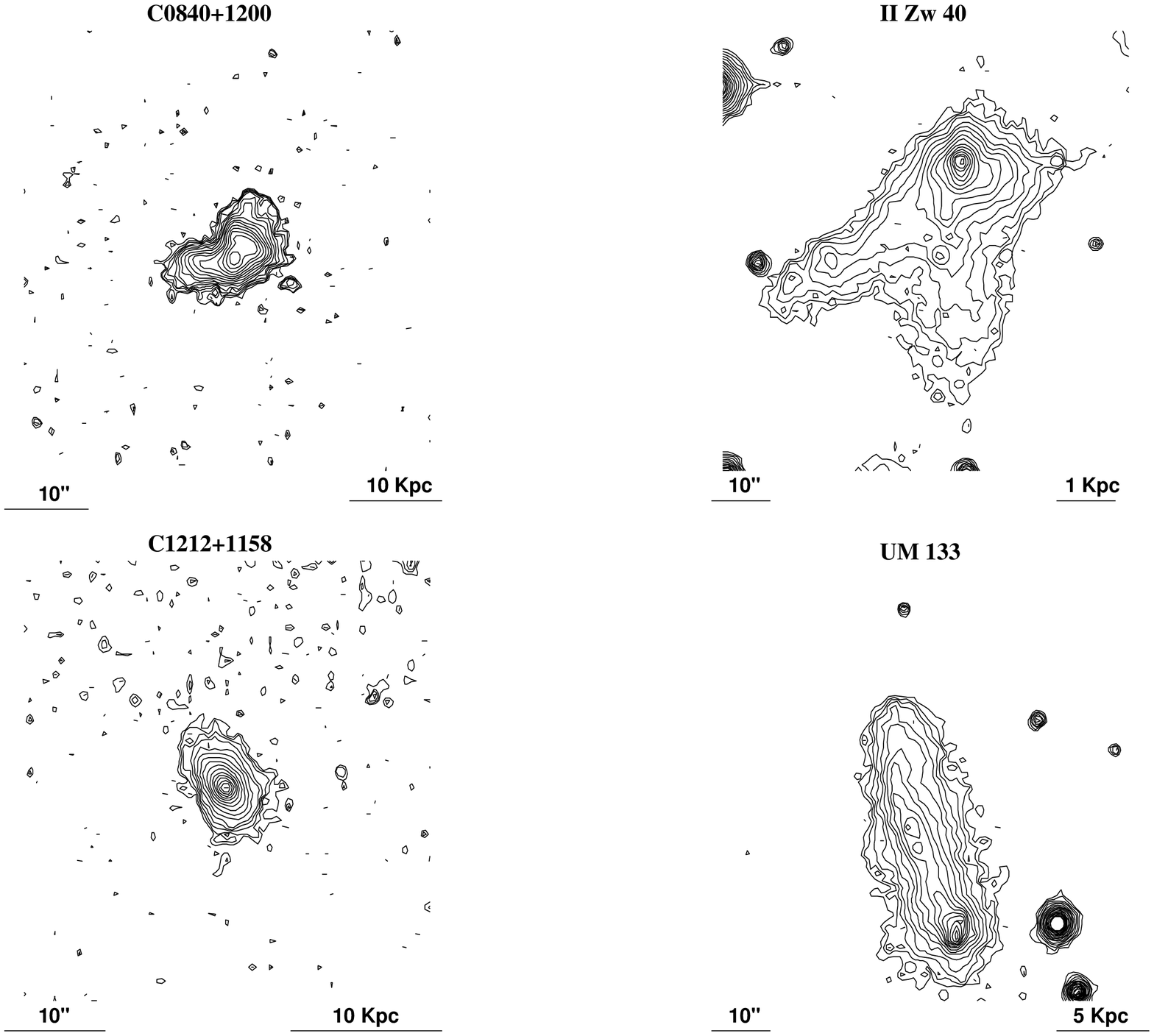}
}
\caption[Contour Maps in logarithmic scale in the V band.]{Contour Maps in
logarithmic scale in the V band. Outer most contours correspond to
V$\sim 25~ mag/{\tt ''}^2$.  North is top and East is left.  Galaxy
name is shown above each map. Scales are shown below each
contour map.}
\label{fig:allcont}
\end{figure*}

\addtocounter{figure}{-1}
\begin{figure*}
\protect\centerline{
\epsfxsize=5.5in\epsffile[ 1 1 710 570]{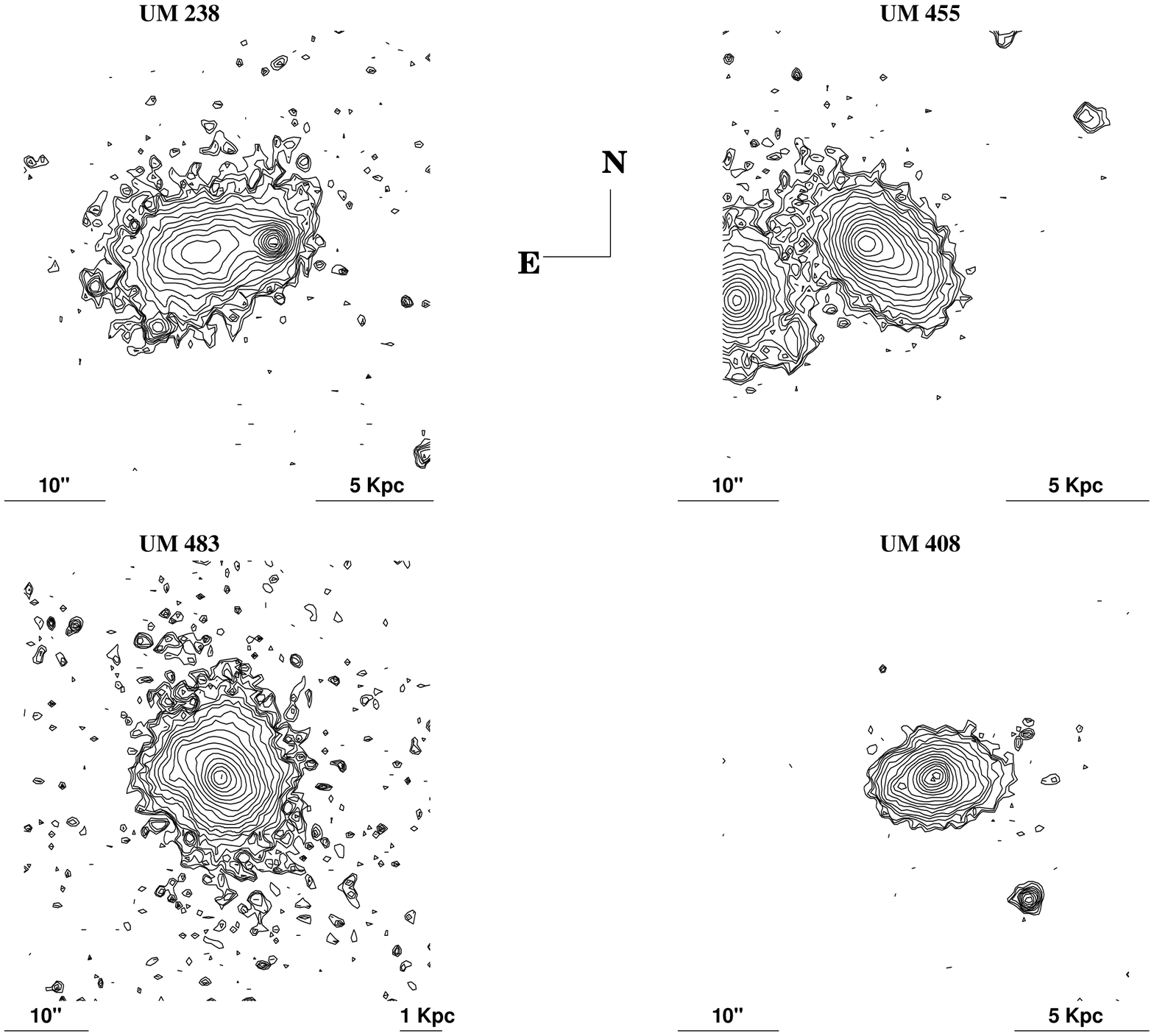}
}
\protect\centerline{
\epsfxsize=5.5in\epsffile[ 1 1 710 570]{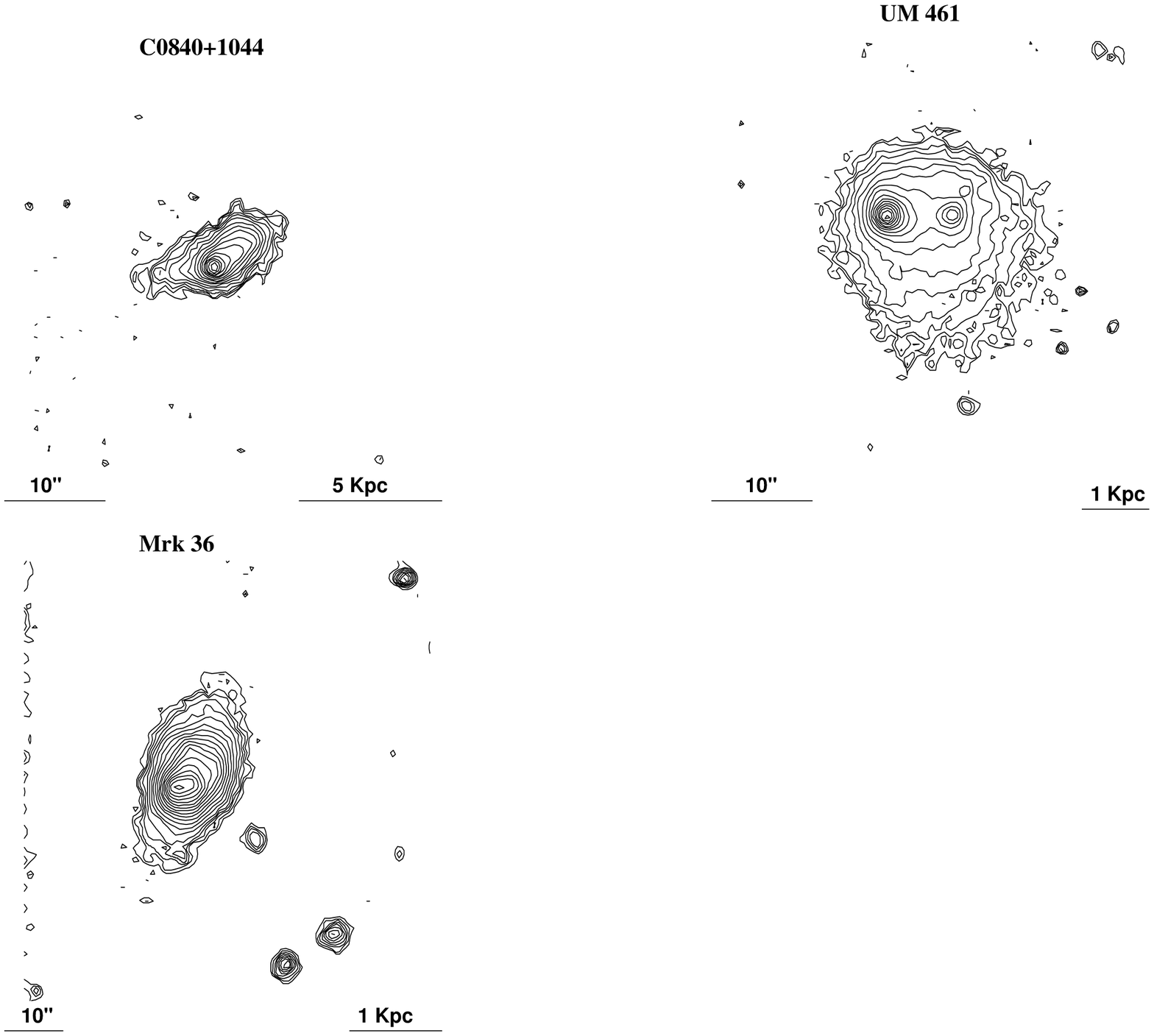}
}
\caption{continue}
\end{figure*}

Figure~\ref{fig:allcont} shows the contour maps in the V bandpass of
the objects in the sample.  The objects are shown from top left to
bottom right in decreasing order of intrinsic luminosity. The contours
are slightly smoothed, typically block-averaged by 3x3 pixels.  The
outer most contours correspond to V$\sim$ 25 mag/${\tt ''}^2$. The
orientation and scales are shown in the figures.  We describe below
some of the basic morphological features for the galaxies in the
present sample.  The morphology for some of these objects have also
been described in Telles (1995) and Telles, Melnick \& Terlevich
(1996).  These are marked with an $\ast$ after the name.  

\begin{description}

\item[C08-28A$\ast$] 0842+162 (Mrk 702 = C0843+1617) \\ 
This is the highest luminosity galaxy in the sample. Its morphology is
very irregular showing 3 individual gigantic HII region complexes
along the E-W direction.  Apparent \,``fans\,'' are visible emerging
from the largest starburst (West-most blob) in both North and South
directions.  The fainter southern extension is of very low surface
brightness and is not clearly visible in this contour map.
With M$_{\rm V} = -22.22$, this Markarian galaxy is by no means a dwarf system.

\vspace{2mm}

\item[UM 167] (NGC 7741) \\ This galaxy is an archetype of the
Starburst Nucleus galaxies (Weedman \etal 1981).  The contour map of
this starburst galaxy is affected by a bright star in the
South. Nevertheless, one can clearly see this interacting pair which
consists of two spiral galaxies.  They appear to be bridged by their
arms from the East of UM167.  The companion of UM167 does not show
emission lines in the narrow band images.  The nuclear starburst in
UM167 is very intense and compact.  It also has circumnuclear activity
regions.  A detailed spectroscopic study of this starburst galaxy has
recently been made by the GEFE collaborators (Gonz\'alez-Delgado \etal
1995; Garc\'{\i}a-Vargas \etal 1996).

\vspace{2mm}

\item[UM 448$\ast$] (Mrk 1304, ARP 161) \\
This peculiar starburst galaxy has a large \,``fan\,'' extending over
1${\tt '}$ to the South-West direction from the center of the large
line emitting region.  No other tail or fan-like extensions is seen on
the opposite side.  A \,``blow-up\,'' of the inner structure seems to
reveal double structure with apparent spiral shapes.  It is worth
noting that only the inner structure is seen on the narrow band
images. There is not strong gaseous line emission in the large
extension in this object.

\vspace{2mm}

\item[C1409+1200] This is the largest redshift object in this sample
(z=0.056).  Its somewhat regular appearance may well be misleading due
to a resolution effect.  In detail a knot can be seen to the
South-East direction which may be associated with this galaxy.  Faint
extensions are seen to the North-West and South-West directions as
well.  These may be the \,``tip of the iceberg\,'' of large extensions
and irregular morphology.

\vspace{2mm}

\item[C0840+1200$\ast$] This galaxy has a peculiar L shape morphology.
It may represent a double structure.  Low spatial resolution and S/N due to
the moderate redshift (z=0.030) of this object is likely to hamper our
detection of some faint details of its outer structure.

\vspace{2mm}

\item[II Zw 40$\ast$] (UGCA 116) \\ 
Together with IZw18, this galaxy is one of the prototypes of HII
galaxies (Sargent \& Searle 1970).  II Zw 40 is a nearby object with a
compact core and double structure.  Fan extensions extend to the South
and South-East suggesting a system undergoing merging of two galaxies.
This object seems to represent a borderline from the more luminous
disturbed objects to less luminous regular galaxies.

\vspace{2mm}

\item[C1212+1148] Regular object with one main starburst region 
in its center.

\vspace{2mm}

\item[UM 133$\ast$] (MCG1-5-30). \\ 
This object has regular outer isophotes but its main burst lies on the
South-most position in the galaxy.  Smaller and much weaker line
emitting regions are also seen along the main body in the narrow band
image.

\vspace{2mm}

\item[UM 238]  Analogous to UM133, this object has a very strong compact star
forming region in one end (West) of the main body.  The outer isophotes
are also regular with no sign of distorted fans or tails.

\vspace{2mm}

\item[UM 455$\ast$] The line emitting region is not centered in relation to
the regular shape of its outer isophotes.  It lies slightly to the North-East
direction.

\vspace{2mm}

\item[UM 483] (Mrk 1313) \\
This object is a nearby low luminosity galaxy with weak star formation
activity in the center of its regular structure.  It does not have
apparent spiral structures or disturbed morphology as it is common
case in more luminous Markarian starburst galaxies.

\vspace{2mm}

\item[UM 408]  Elliptical regular shape, centered burst.

\vspace{2mm}

\item[C0840+1044$\ast$] This object has an apparent extension in the East
direction in an otherwise regular morphology.  Its burst lies on one end
of the regular structure.

\vspace{2mm}

\item[UM 461$\ast$]   This is a double burst object embedded in regular
outer elliptical isophotes.  The double structure is displaced in relation
to the outer structure.  The eastern knot (UM461A) is a very strong and
compact line emitting region.

\vspace{2mm}

\item[Mrk 36] (Haro 4). \\ 
This is the lowest redshift object in the sample.  It has elliptical outer
isophotes and its burst is located in the South-East part of the galaxy.

\end{description}

\subsection{The Sizes of the star forming regions}\label{stellar:sizes}

\begin{table}
\centering
\caption{Nebular Sizes.}
\label{tab: burst sizes}
\begin{tabular}{lcrr} \hline
{\em object} & seeing & \multicolumn{2}{c}{Nebula Radius} \\
            &  $(\,'')$ &  $(\,'')$ & (pc) \\ \hline
%
UM238        &   0.7 &   2.5 &   1021 \\ 
UM133        &   0.9 &   4.1 &   1083 \\ 
UM408        &   0.9 &   2.5 &    883 \\ 
IIZw40       &   1.0 &   9.4 &    825 \\ 
C0840+1040   &   0.7 &   1.6 &    553 \\ 
C0840+1201   &   1.2 &   4.4 &   3835 \\ 
C08-28A    &   1.6 &   8.1 &  12662 \\ 
Mark36       &   1.7 &   6.6 &    384 \\ 
UM448        &   1.0 &   7.9 &   4122 \\ 
UM455        &   1.4 &   3.9 &   1362 \\ 
UM461        &   0.9 &   5.2 &    751 \\ 
UM483        &   1.2 &   5.5 &   1125 \\ 
C1212+1148   &   0.9 &   2.7 &   1821 \\ 
C1409+1200   &   1.2 &   2.0 &   3308 \\ 
UM167        &   1.2 &  12.6 &   3298 \\ 
 \hline
\end{tabular}
\end{table}

We have determined the radius of the ionized gas regions from the
narrow band images, as described in Section~\ref{stellar:reduction}.
Table~\ref{tab: burst sizes} lists the Gaussian FWHM mean seeing for
the observations of each galaxy as well as the nebular mean radius
(R$_{neb} = \sqrt{\frac{Area~of~nebula}{\pi}}$) in arcseconds and in
parsec.  It can be seen that the sizes of the ionized regions in HII
galaxies range from several hundred parsec in the low luminosity
objects to a few Kpc in the most luminous objects.

\begin{figure}
\protect\centerline{
\epsfxsize=3in\epsffile[ 20 150 590 720]{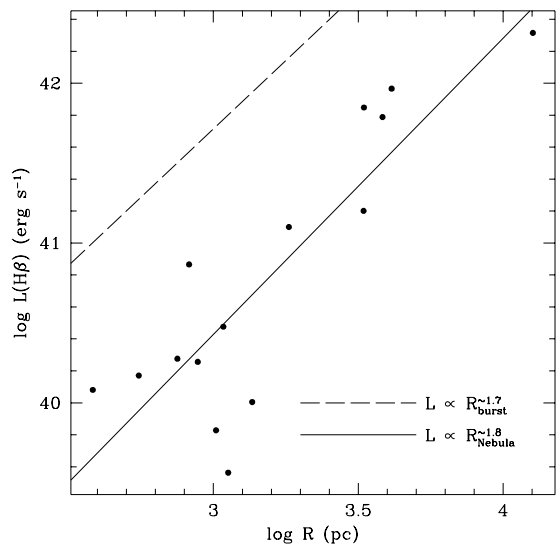}
}

\caption[L(H$\beta$) - linear size relation.]{L(H$\beta$) - linear size
relation.  The solid line is the least square fit to the L(H$\beta$)
-- Nebula size relation.  The dashed line is the L(H$\beta$) -- Burst
size relation as derived in TT93 (see text).}
\label{fig: l-r} 
\end{figure}

\begin{table}
\centering
\caption[Characteristic Dimensions of 30 Doradus in the
LMC.]{Characteristic Dimensions of 30 Doradus in the LMC.  This table
is reproduced from table 1 of Walborn (1991).}
\label{tab:30Dor}
\begin{tabular}{lrl} \hline
        & \multicolumn{2}{c}{Linear Diameters} \\ \hline
LMC                 & 5 & Kpc \\
30 Doradus Region   & 1 & Kpc \\
30 Doradus Nebula   & 200 & pc \\
30 Doradus Cluster  & 40 & pc \\
R136                & 2.5 & pc \\
R136a               & 0.25 & pc \\ \hline
\end{tabular}
\end{table}

Figure~\ref{fig: l-r} shows the [L(H$\beta$) -- R$_{neb}$] relation
from the data in Tables~\ref{stellar:positions} and~\ref{tab: burst
sizes}.  The solid line is a least-square fit to the data.  Also shown
as a dashed line is the [L(H$\beta$) -- R$_{burst}$] as determined in
Telles (1995) and Telles \& Terlevich (1993, TT93).  As we pointed out
in TT93, R$_{burst}$ is only a scaled version of the
\,``true\,'' core radius of the ionizing stellar cluster, but we did
not attempt to investigate what R$_{burst}$ in fact represents.  Here,
from figure~\ref{fig: l-r}, we note that, for the same luminosity
range, the slope of the [L(H$\beta$) -- R$_{neb}$] relation is the
same as that of the [L(H$\beta$) -- R$_{burst}$] relation from TT93.
However, R$_{burst}$ is 0.7 dex (5 $\times$) smaller than the nebula
size measured here.  If we combine this information with the size
scales for 30 Doradus, the nearest giant HII region (in the Large
Magellanic Cloud) for which high resolution observations are available
(see Walborn 1991 and references therein) we may, by analogy, have an
indication of what R$_{burst}$ in TT93 is measuring.
Table~\ref{tab:30Dor} reproduces Table 1 of Walborn (1991) which gives
the linear sizes of 30 Doradus region and its contents.  It is
interesting to note that the size of the ionizing cluster (40 pc)
defined as the distribution of all O and B stars is 5 times smaller
than the size of the 30 Dor nebula (200 pc).  Therefore, we can assume
that the R$_{burst}$ in TT93 represents the total size of the stellar
cluster (as defined above) for the HII galaxies as well.  However, the
size of the dense core of the ionizing cluster, where the bulk of the
total UV luminosity is generated, is more than one order of magnitude
smaller than R$_{burst}$.  This confirms the earlier expectation that
the \,``true\,'' core of the ionizing cluster in HII galaxies to be of
the order of a few parsec across at most and will remain unresolved in
ground-based observations.  Vacca (1994) and Conti \& Vacca (1994)
observed a subset of HII galaxies whose integrated spectra exhibit
broad stellar emission lines due to the presence of hundreds to
thousands of Wolf-Rayet stars (Wolf-Rayet galaxies) using the Faint
Object Camera aboard the Hubble space Telescope in the UV light.
Large star-forming regions which appear to be single units in the
optical images are resolved into numerous discrete compact bright
knots.  They find that these starburst knots are typically less than
100 pc in size and generally too small and closely spaced to be
detected individually in the ground-based optical images.  However,
the knots contain large numbers of hot (O, B, and Wolf-Rayet) stars
and are typically several times more luminous than 30 Doradus.  For
these reasons, in order to have a better understanding of the
ionization structure in HII galaxies, it is imperative to have high
spatial resolution observations.  This makes the lowest redshift HII
galaxies good targets for further projects with the Hubble Space
Telescope.

\subsection{Luminosity profiles}

\begin{table*}
\centering
\caption{Exponential fitting parameters.}
\label{tab:fitresults}
\begin{tabular}{lcrcrcc} \hline
{\em object} & profile & $r_0$ & $\mu_0$ & $r_{eff}^{prof}$ & $\mu_{eff}^{prof}$ & $I_{extrap}$ \\
             & type &$(\,'')$ &  mag/$_{\,''}~^2$ & $(\,'')$ &  mag/$_{\,''}~^2$ & mag  \\ \hline
UM238      & dd & 4.53 & 20.95 &  7.60 & 22.82 & 15.67 \\
UM133      & dd & 7.95 & 20.23 & 13.34 & 22.10 & 13.73 \\
UM408      & d  & 1.54 & 19.70 &  2.58 & 21.57 & 16.77 \\
IIZw40     & bd &13.26 & 20.64 & 22.25 & 22.51 & 13.03 \\
C0840+1044 & d  & 1.84 & 20.73 &  3.08 & 22.60 & 17.41 \\
C0840+1201 & d: & 1.77 & 19.29 &  2.97 & 21.17 & 16.06 \\
C08-28   & bd & 9.23 & 22.01 & 15.48 & 23.89 & 15.19 \\
Mark36     & d  & 3.18 & 19.58 &  5.33 & 21.45 & 15.07 \\
UM448      & bd & 3.67 & 18.78 &  6.16 & 20.65 & 13.96 \\
UM455(R)   & d  & 2.28 & 20.45 &  3.82 & 22.33 & 16.67 \\
UM461      & dd & 3.07 & 20.03 &  5.16 & 21.90 & 15.60 \\
UM483      & d  & 2.08 & 18.81 &  3.50 & 20.68 & 15.22 \\
C1212+1148 & d  & 1.09 & 19.09 &  1.83 & 20.97 & 16.91 \\
C1409+1200 & dd:& 0.74 & 18.91 &  1.25 & 20.78 & 17.55 \\
UM167      & bd &15.63 & 20.26 & 26.22 & 22.13 & 12.30 \\
\hline
\end{tabular}
\end{table*}

\begin{figure*}
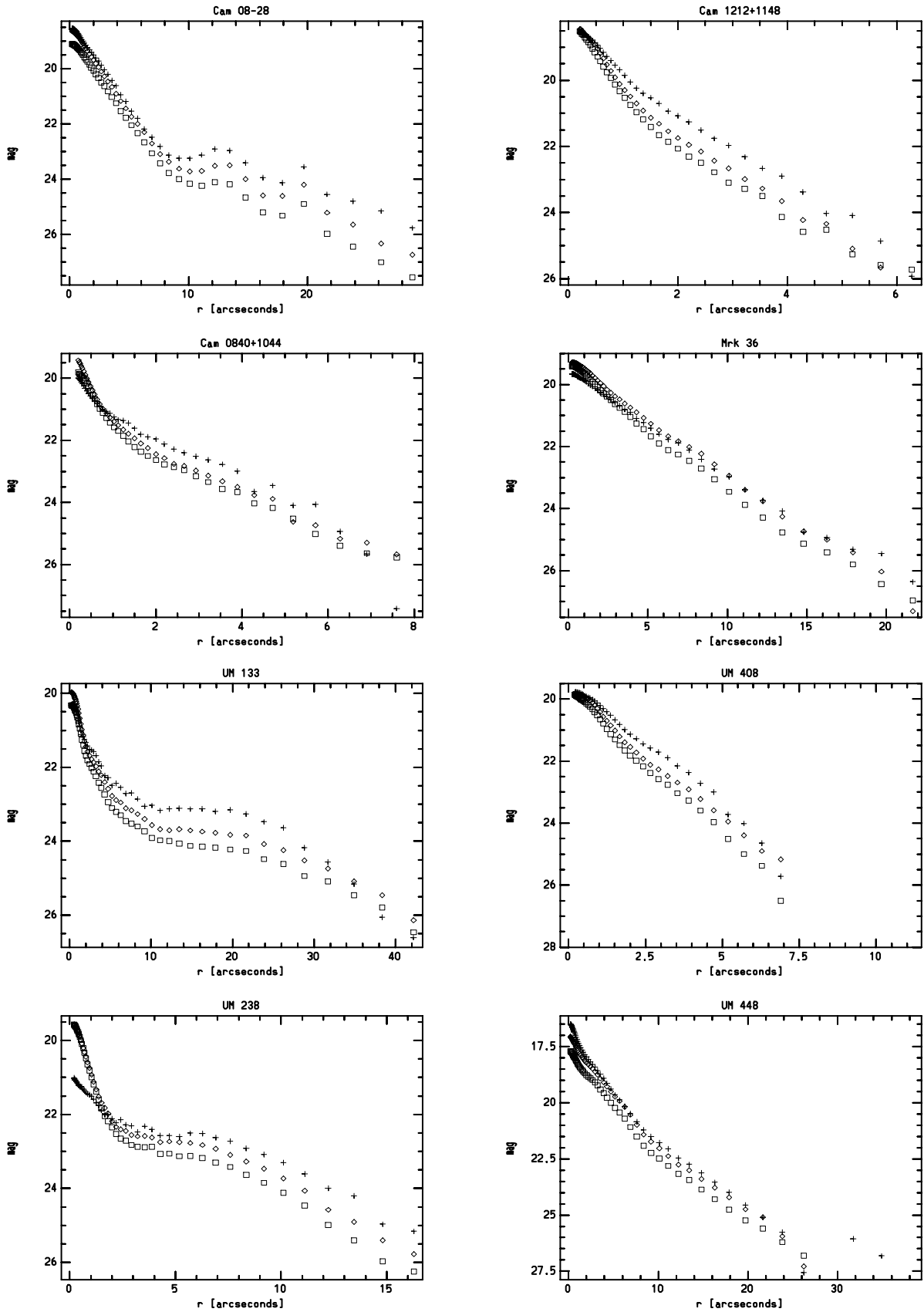
 
\protect\centerline{
\epsfxsize=6.8in\epsffile[ 70 240 540 550]{chap_4.profa}
}
\protect\centerline{
\epsfxsize=6.8in\epsffile[ 70 240 540 550]{chap_4.profb}
}
\caption{Luminosity profiles in the V (box), R (diamond) and I (plus) 
bandpass.}
\label{fig:allprof}
\end{figure*}

We favour the simpler azimuthally averaged mean profiles over ellipse
fitting algorithms to obtain the luminosity profiles from
two-dimensional photometry for these systems with irregular isophotes.
The conclusion from an exercise is that the ellipse fitting routines
only introduces parameters that do not provide additional information
on the underlying brightness distribution of HII galaxies, often
giving results affected by the irregularities of the isophotes in the
star-forming regions in these objects.  The exponential law was
confirmed to represent well the outer, parts of the profiles.  We are
interested in assessing the properties of the underlying galaxy in HII
galaxies and possible structural links with other types of dwarf
galaxies. Therefore, we derive the luminosity profiles in the surface
brightness representation for the three broad band filters, as shown
in figure~\ref{fig:allprof}.  In this form the exponential scaling law
[$\mu(r) = \mu_0 + 1.086 \frac{r}{r_0}$] becomes a straight line and
the only two shape-free parameters from a fit to the profile are:
$\mu_0$, the central surface brightness, and $r_0$ the exponential
scale length.  We fitted this exponential scaling law to the
circularly averaged surface brightness profiles in the I filter only.
The I filter is not affected by the line emission from the gaseous
component at a significant level. Therefore, the irregularities
(i.e. bumps) should be smaller and hence profiles are expected to be
smoother.  The light from the ionizing OB stars will contribute little
to the total fluxes at these wavelengths.  The resulting I profiles
will, thus, reflect the continuum contribution of the cooler stellar
population, possibly providing a handle on the underlying galaxy.
Figure~\ref{fig:allfits} shows the resulting exponential fits to the
outer parts of the I profiles in each galaxy.  The range of radii at
which the profile is fitted is rather arbitrary.  We fitted the
profiles to surface brightness levels as low as where the local S/N
was $\approx$ 1 and as high as the outer profile trend. The solid
horizontal line along the x axis represents the range of the profile
fitted.  Table~\ref{tab:fitresults} presents the results from these
fits.  Column 1 gives the name of the galaxy. Column 2 the profile
type as discussed in Telles (1995) and Telles, Melnick \& Terlevich
(1996).  Columns 3 to 7 give the scale length ($r_0$ in arcseconds),
the central surface brightness ($\mu_0$), the effective radius
($r_{eff}^{prof}$), the effective surface brightness
($\mu_{eff}^{prof}$) and the extrapolated total I magnitude
($I_{extrap}$), respectively. These quantities are all derived from
the fitting of an exponential scaling law to the outer parts of the I
profiles, except in the case of UM455 which are the results of the fit
to the R profile.

\addtocounter{figure}{-1}
\begin{figure*} 
\protect\centerline{
\epsfxsize=6.8in\epsffile[ 70 240 540 550]{chap_4.profc}
}
\protect\centerline{
\epsfxsize=6.8in\epsffile[ 70 240 540 550]{chap_4.profd}
}
\caption{continue}
\end{figure*}

\begin{figure*}
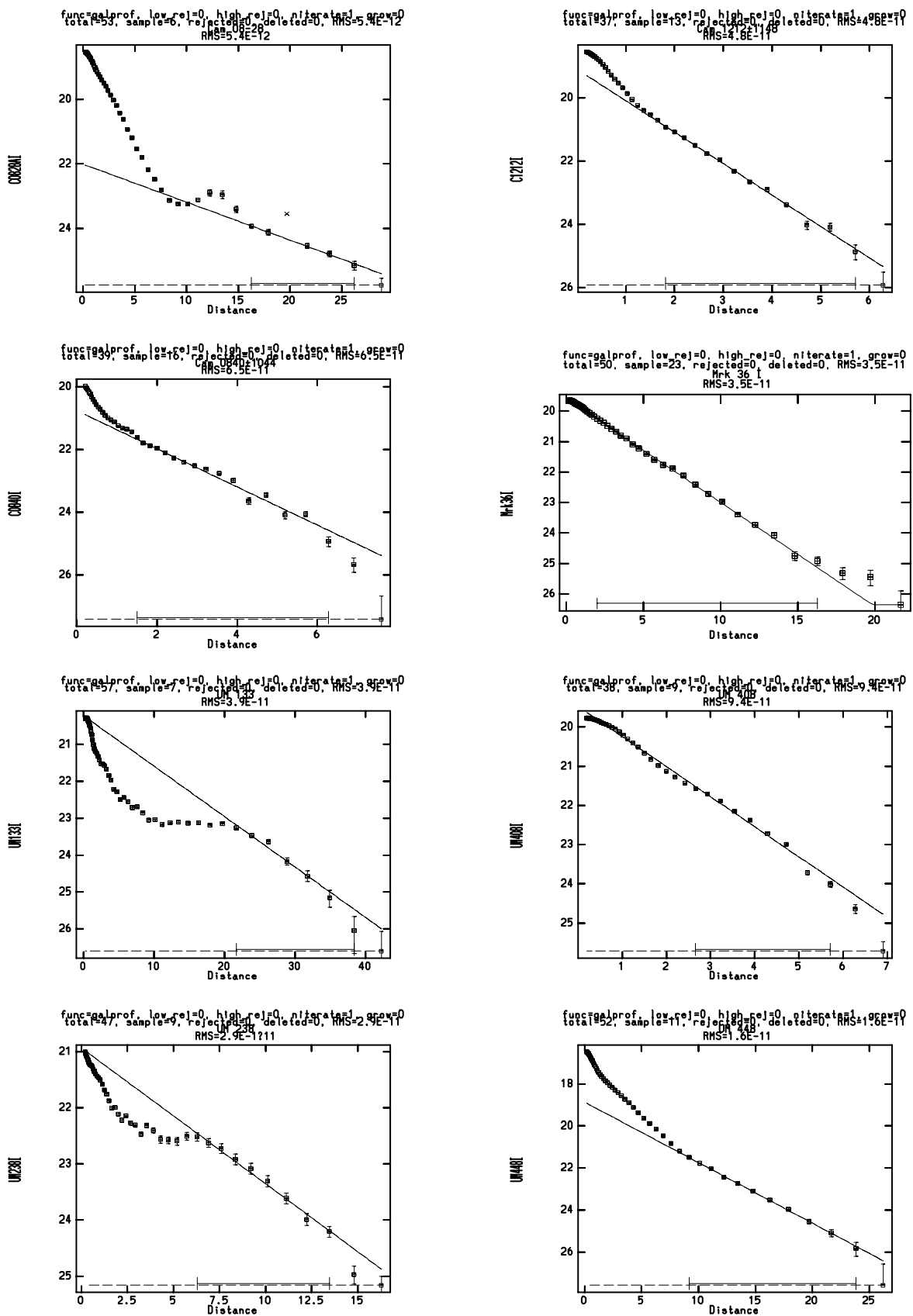
 
\protect\centerline{
\epsfxsize=6.8in\epsffile[ 70 240 540 550]{chap_4.fita}
}
\protect\centerline{
\epsfxsize=6.8in\epsffile[ 70 240 540 550]{chap_4.fitb}
}
\caption[Exponential fit to the I bandpass luminosity profiles.]{Exponential
fit to the I bandpass luminosity profiles. The x axis is the radius in
arcseconds.  The y axis is in I mag/$_{\,''}~^2$.}
\label{fig:allfits}
\end{figure*}

\addtocounter{figure}{-1}
\begin{figure*} 
\protect\centerline{
\epsfxsize=6.8in\epsffile[ 70 240 540 550]{chap_4.fitc}
}
\protect\centerline{
\epsfxsize=6.8in\epsffile[ 70 240 540 550]{chap_4.fitd}
}
\caption{continue}
\end{figure*}

From the colour profiles, the most general result that can be drawn is
that these systems are composed of two components: a constant colour
component in the outer regions (the extensions) and a colour gradient
component (the burst).  The latter shows that the colours generally,
but not always, become bluer in V--I in the inner parts (UM 448 is a
counter example).

\begin{table*}
\centering
\caption[HII Galaxies extinction corrected colours.]{
HII Galaxies extinction corrected colours: Total colours, Starburst
colours corrected for extinction using C(H$\beta$) and Starburst
colours corrected for the contribution of emission lines.  Total
colours are corrected for galactic extinction and include emission
lines.  Both starburst colours have also been corrected for the
contribution of the underlying galaxy (see text for more details). }
\label{tab: colours}
\begin{tabular}{lccccccccc} \hline
{\em object} &\multicolumn{3}{c}{Total}
             &\multicolumn{3}{c}{Starburst$^{corrected}$}
             &\multicolumn{3}{c}{Starburst$_{continuum}^{corrected}$} \\ \hline
 &  V--R & V--I & R--I  & V--R & V--I & R--I  & V--R & V--I & R--I \\ \hline
UM238        & 0.35 & 0.63 & 0.28 &  0.09 & -0.95 & -1.04 &  0.39 & -0.14 & -0.54  \\
UM133        & 0.35 & 0.78 & 0.43 &  0.29 &  0.09 & -0.21 &  0.20 &  0.39 &  0.19   \\
UM408        & 0.25 & 0.65 & 0.40 &  0.21 &  0.35 &  0.14 &  0.27 &  0.62 &  0.35   \\
IIZw40       & 0.25 & 0.02 &-0.23 &  0.84 &  0.39 & -0.45 &  1.04 &  1.20 &  0.17   \\
C0840+1040   & 0.33 & 0.63 & 0.29 &  0.21 & -0.11 & -0.32 &  0.16 &  0.11 & -0.05  \\
C0840+1201   & 0.27 & 0.26 &-0.01 &  0.29 &  0.07 & -0.22 &  0.15 &  0.17 &  0.02   \\
C08-28A    & 0.42 & 0.73 & 0.31 &  0.41 &  0.39 & -0.02 &  0.33 &  0.53 &  0.20   \\
Mark36       & 0.35 & 0.21 &-0.13 &  0.29 & -0.08 & -0.36 &  0.09 &  0.26 &  0.17  \\
UM448        & 0.53 & 0.72 & 0.19 &  0.54 &  0.72 &  0.18 &  0.44 &  0.86 &  0.42   \\
UM455        & 0.28 & 0.51 & 0.24 &  0.30 & -0.02 & -0.32 &  0.38 &  0.30 & -0.08  \\
UM461        & 0.21 & 0.44 & 0.23 &  0.19 & -0.23 & -0.42 &  0.35 &  0.44 &  0.09   \\
UM483        & 0.30 & 0.56 & 0.26 &  0.32 &  0.55 &  0.23 &  0.27 &  0.57 &  0.30  \\
C1212+1148   & 0.24 & 0.71 & 0.46 &  0.22 &  0.42 &  0.20 &  0.24 &  0.74 &  0.50   \\
C1409+1200   & 0.08 & 0.13 & 0.05 & -0.03 & -0.24 & -0.21 & -0.17 & -0.06 &  0.11  \\
UM167        & 0.31 & 0.52 & 0.22 &  0.43 &  0.78 &  0.35 &  0.42 &  0.78 &  0.36  \\\hline
\end{tabular}
\end{table*}

\subsection{The colours of the starburst and of the underlying galaxy}
\label{stellar:burst_galaxy_col}

From what follows, we will treat the study of the stellar populations
in HII galaxies as two separate components:
 
\begin{description} 
\item[{\em The Starburst}] The region where gas emission was detected through 
narrow band observations.  The stellar component within these regions
is dominated by the young ionizing stars formed in the present episode
of star formation.  Once corrected by the effect of line emission and
for the contribution of the underlying galaxy within this region, we
attempt to have a handle on the properties of the ionizing stellar
cluster.

\vspace{2mm}
 
\item[{\em The underlying galaxy}] The extensions are the outer regions of 
the galaxies which are not dominated by line emission from the
starburst. We assume that the light from these regions are
representative of a stellar component formed previously. Thus, we
attempt to have a handle on the properties of the underlying galaxy.
 
\end{description} 

We have corrected the broad band magnitudes for the line emission in
order to have a better insight on the colours of the ionizing clusters
embedded in the nebula regions.  This correction is described in
Salzer, MacAlpine \& Boroson (1989) and derived in appendix A for the
present purpose.  Table~\ref{tab: colours} shows the V R I total (whole
galaxy) colours corrected for galactic extinction only, as described
in section~\ref{stellar:integrated}. Table~\ref{tab: colours} also
presents the results for the colours of the ionized region
(Starburst$^{corrected}$) corrected for total external extinction
derived from the Balmer Decrement [C(H$\beta$)] from SCHG and assuming
Case B recombination.  The colours of the starburst are also corrected
for the contribution of the underlying galaxy by assuming the mean
surface brightness of the extensions to be constant and representative
of the underlying galaxy within the starburst regions.  The last three
columns of Table~\ref{tab: colours} give the colours of the ionizing
stellar cluster with an additional subtraction of the contribution of the
emission lines (Starburst$_{continuum}^{corrected}$), as derived in
appendix A.

\subsubsection{Properties of the underlying galaxy}

Table~\ref{tab:extcolours} shows the measurements of the colours and
mean surface brightness of the extensions.  The here-called
extensions, derived directly from the CCD frames for each object, is
the region beyond the starburst and 2 S/N above sky noise.  It
represents the regions in the galaxy where gas emission is
non-existent or faint, and are representative of the photometric
properties of the underlying galaxy in HII galaxies.

\section{Discussion}\label{stellar:discuss}

\subsection{Morphology}

Despite the mixed bag of patterns in the position and multiplicity of
the starburst regions, we may raise some general points found here:

\begin{enumerate}
\item Galaxies with disturbed morphologies often showing extended \,``fans\,''
or tails, irregular or double outer structures also have higher
luminosity.
\item Galaxies with regular outer structure (isophotes) tend to fall
in the lower luminosity regime.
\item Often the starburst is not centered in relation with the outer
isophotes.
\item Double knot objects (such as UM461) seem to be common.
\item Very compact, regular objects with centered starburst are also present.
\end{enumerate}

The results presented here agree with what has been found by Telles,
Melnick \& Terlevich (1996) in the analysis of a sample of 39 galaxies
in the V band, namely that HII galaxies fall into two broad types
based on their overall appearance.  We called Type I, galaxies with
distorted outer isophotes while Type II HII galaxies are regular and
compact.  From figure~\ref{fig:allcont} and the photometry, we confirm
the result that Type I HII galaxies have higher luminosities and
larger emission line widths while as the luminosity decreases we find
galaxies which fall in the Type II classification (regular isophotes,
no signs of disturbed morphology).

These points on the morphology of these objects give rise to
some questions about the processes which may have played an important
role in initiating the starburst.  For instance:

\begin{itemize}

\item Is the disturbed morphology with extended outer structures 
related to colliding or merging systems (e.g. C0828, IIZw40, etc.)?
\item Can double knot (e.g. UM461) mean two gas rich systems in 
process of merging?
\item Are  starbursts  located at one end of a galaxy of high
ellipticity (e.g. UM133, UM238) triggered by the
accumulation of gas in a rotating bar (see Elmegreen 1992)?
\item How does the onset of star formation take place in the most
compact, regular and isolated HII galaxies? (see also Telles
\& Terlevich 1995)

\end{itemize}

All these questions raised by the morphological study of HII galaxies may
help us classify and better understand the fundamental processes ruling
large scale star formation in galaxies.

\begin{table}
\centering
\caption{Colours of the underlying galaxy or extensions and their mean
surface brightness.}
\label{tab:extcolours}
\begin{tabular}{lcccccc} \hline
{\em object} &\multicolumn{3}{c}{Extensions}
             &\multicolumn{3}{c}{$<\mu>_{extensions}$}\\ \hline
 &  V--R & V--I & R--I  & V & R & I  \\ \hline
UM238        &  0.34 &  0.81 &  0.47 & 22.75 & 22.54 & 21.83 \\ 
UM133        &  0.34 &  0.98 &  0.64 & 23.27 & 22.96 & 22.16 \\ 
UM408        &  0.28 &  0.90 &  0.62 & 23.25 & 23.01 & 22.16 \\ 
IIZw40       &  0.24 &  0.45 &  0.21 & 21.55 & 21.58 & 21.24 \\ 
C0840+1040   &  0.24 &  0.75 &  0.51 & 22.99 & 22.68 & 21.99 \\ 
C0840+1201   &  0.23 &  0.55 &  0.31 & 22.79 & 22.81 & 22.42 \\ 
C08-28A    &  0.36 &  1.09 &  0.73 & 23.28 & 23.05 & 22.13 \\ 
Mark36       &  0.46 &  0.72 &  0.26 & 23.26 & 22.91 & 22.25 \\ 
UM448        &  0.44 &  0.66 &  0.21 & 22.98 & 22.65 & 21.91 \\ 
UM455        &  0.24 &  0.78 &  0.54 & 23.01 & 22.81 & 22.16 \\ 
UM461        &  0.21 &  0.73 &  0.52 & 23.41 & 23.25 & 22.27 \\ 
UM483        &  0.25 &  0.62 &  0.37 & 23.25 & 23.16 & 22.49 \\ 
C1212+1148   &  0.27 &  1.20 &  0.93 & 23.26 & 22.91 & 21.89 \\ 
C1409+1200   &  0.23 &  0.74 &  0.51 & 23.35 & 23.26 & 22.52 \\ 
UM167        &  0.38 &  0.45 &  0.07 & 22.71 & 22.05 & 21.88 \\ 
\hline
\end{tabular}
\end{table}

\subsection{Luminosity profiles}

The different types of profiles reflect the different morphologies
among the objects in this sample.  An exponential scaling law is
generally a good fit to the \underline{outer} parts (extensions) of the
I luminosity profiles of HII galaxies.  Remarkable cases where single
exponential fit seem to represent well the whole range of radii of the
profile are: Mrk36, UM408, C1212+1148, C0840+1044, UM455 and UM 483.
In many of these the fit extends to 5 magnitudes in surface
brightness.  For these the integral luminosity of the exponential
\,``disk\,'' and the total fluxes within synthetic circular apertures
are identical within a few per cent.

\begin{figure}
\protect\centerline{
\epsfxsize=3in\epsffile[ 18 150 590 720]{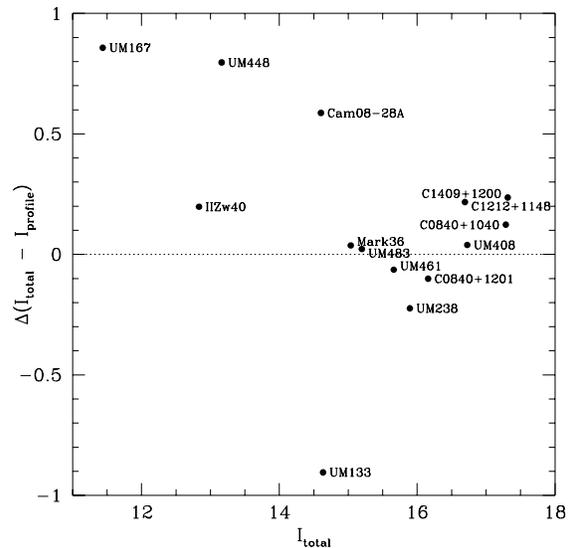}
}
\caption{Difference between the curve of growth ($I_{total}$)
and extrapolated profile fit total magnitude ($I_{profile}$).}
\label{fig:Iprof_I}
\end{figure}

The goodness of an exponential law to represent the light distribution
of a galaxy can be tested when one compares the model free parameters
($m_{tot}$, $m_{eff}$, $r_{eff}$ and $<\mu>_{eff}$), measured from the
curve of growth, with the same parameters derived from the
extrapolation of the fit to their luminosity profile.  For instance,
figure~\ref{fig:Iprof_I} shows a plot of the residual between total
magnitude measured from the asymptotic value from the curve of growth
($I_{total}$) and the extrapolated total magnitude derived from the
integration of the derived exponential law with its respective fit
parameters to infinity ($I_{profile}$) against $I_{total}$.  The
exponential fit to the extensions best represents the total light
distribution for galaxies near the dashed line.

For the large luminous galaxies in the sample their profiles seem to
present a composite behaviour with a steeper excess light in its inner
part, corresponding to the burst region, in relation with a more
extended flatter outer parts.  Such is the case for C0828, UM448,
UM167 and possibly IIZw40.  The total fluxes derived from the
extrapolation of the exponential fit to the large radius range will,
thus, underestimate the total galaxy luminosity.  Galaxies with a
double burst morphology have a \,``platform\,'' or \,``bump\,'' in
that profile due to the secondary burst component.  This occurs
because the profiles are centered on the peak luminosity corresponding
to the strongest burst.  These profiles will show a dip between the
two peaks (see the case for UM461 and UM238) which will make the
extrapolation of the outer fit, in this case, overestimate the total
flux of the galaxy.  It is interesting to note that the stronger burst
in these objects with double morphology have very large equivalent
width of H$\beta$ [W(H$\beta$) $>$ 100] in the sample.  Whether this
is indicative of any basic physical process at play such as dynamical
evolution or merging is not yet known.  UM 133 also shows a similar
platform type of profile but a relatively deeper dip.  As opposed to
the double systems though, this object seems to have a Magellanic-like
or cometary (Loose \& Thuan 1985) morphology.  These are apparently
bar systems in which the giant dominant HII region concentrates in one
extreme of the \,``bar\,''.

\subsection{Comparison of the colours of the underlying galaxies with
other dwarf galaxies}

\begin{figure*}
\protect\centerline{
\epsfxsize=6.0in\epsffile[ 18 150 590 700]{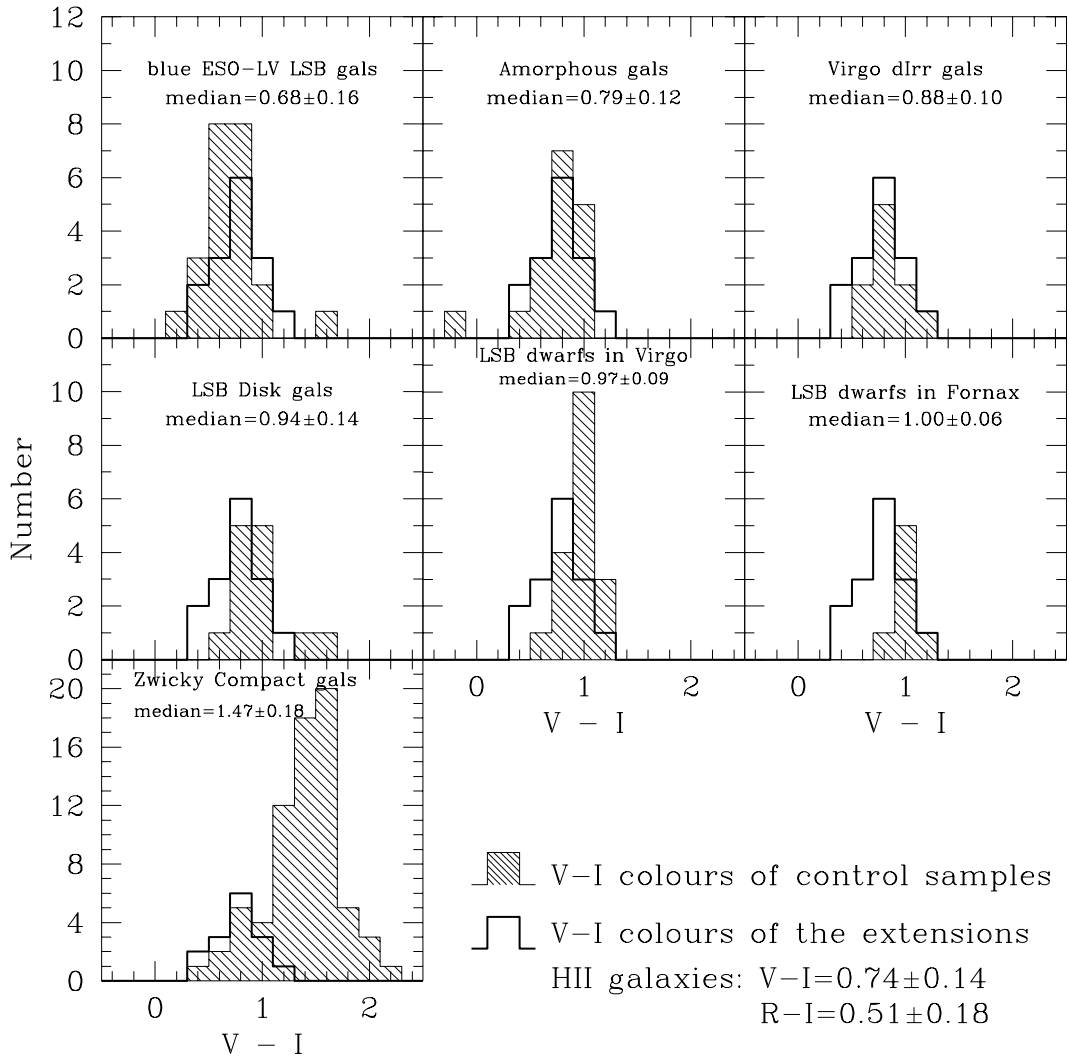}
}
\caption{Comparison of the colours of the {\em extensions} of HII galaxies 
control samples of dwarf galaxies. }
\label{stellar:col_hist}
\end{figure*}

We have used an empirical approach in order to have some insight on the
relation between HII galaxies and quiescent types of dwarf galaxies by
directly comparing the colours of a sample of real dwarf galaxies with
the colours we have derived for the underlying galaxy in the 15 HII
galaxies of our sample (Table~\ref{tab:extcolours}).
 
The colours of the galaxies in the control samples we have used here
have been  converted to the Cousins photometric system
by the use of relations found in Bessel (1979) and
Wade \etal (1979).  The  samples used are:
 
\begin{description} 
 
\item[{\em Blue ESO-LV LSB galaxies}] (R\"{o}nnback \& Bergvall 1994) \\ 
Blue low surface brightness galaxies selected from the ESO-Uppsala
catalogue.  Their blue colour is part of the selection criteria as
well as their low surface brightness ($\mu^{eff}_B > 23.5$ mag
arcsec$^{-2}$).  The result is a sample dominated by irregular or late
type Sd,Sm galaxies.

\vspace{1mm} 

\item[{\em Amorphous galaxies}] (Gallagher \& Hunter 1987) \\ 
Galaxies with E/S0 like morphologies whose other global properties 
resemble irregular galaxies and Magellanic irregulars.  

\vspace{1mm} 

\item[{\em Virgo dIrr galaxies}] (Bothun \etal 1986) \\ 
Dwarf irregular galaxies in the Virgo cluster. 

\vspace{1mm} 
 
\item[{\em LSB disk galaxies}] (McGaugh \& Bothun 1994) \\ 
Low Surface Brightness galaxies ($\mu_0 > 23$ mag arcsec$^{-2}$) 
from the Uppsala General Catalogue and from Schombert \etal (1992). 

\vspace{1mm} 
 
\item[{\em LSB dwarfs in Virgo}] (Impey, Bothun \& Malin 1988) 
 
Low surface brightness galaxies in the Virgo cluster initially 
selected on photographically amplified UK Schmidt plates.  Most of 
these are classified as dwarf elliptical galaxies due to their 
morphology and low HI content. 

\vspace{1mm} 
 
\item[{\em LSB dwarfs in Fornax}] (Bothun, Impey \& Malin 1991) 
 
As above, but for the Fornax cluster of galaxies.

\vspace{1mm} 
 
\item[{\em Zwicky compact galaxies}] (Moles \etal 1987) 
 
This is a heterogeneous set of objects selected basically by their
appearance on the Palomar Sky Survey prints by Zwicky (1964).  The
bluest subset was selected for a spectroscopic work by Sargent (1970).
He found that these galaxies tend to show sharp emission lines in
their spectra.  This is the origin of the name blue compact galaxies
for this type of galaxies. HII galaxies have similar colours to this
blue tail of the distribution of Zwicky galaxies.  The distribution of
colours of the complete sample shows that most Zwicky compact are red
galaxies with colours typical of elliptical galaxies.
 
\end{description}

Figure~\ref{stellar:col_hist} shows the result of the comparison of
the \nobreak{V--I} colours of the {\em extensions} of HII galaxies (the
underlying galaxy) with the total colours of different samples of
dwarf galaxies.  The solid line histograms represent the colours of
HII galaxies while the hatched histograms represent the distribution
of V--I colours of the galaxies in the control samples.  The panels
were arranged in order of decreasing best match to the distribution of
the colours of HII galaxies.  The median and standard deviation of the
median colours for each of the comparison sample are given in each
panel.  For HII galaxy these values are ${\rm V}-{\rm I} =
0.74\pm0.14$ and ${\rm R}-{\rm I} = 0.51\pm0.18$.  It is clear from
the top panels that the underlying galaxy in HII galaxies are best
compared with low luminosity late type galaxies (top panels) as
opposed to early type dwarfs (e.g. LSB dwarfs ellipticals in Virgo or
Fornax).
 
The remarkably blue colours of the underlying galaxy may indicate the
lack of an old diffuse red disk which is associated with an old disk
component such as in high surface brightness spiral galaxies or
extremely low metallicity.

It is interesting to note that the observed distribution of mean
surface brightness of the extensions ($<\mu_{ext}>_{\rm V} \approx
23.0\pm0.5$ mag/${\tt ''}^2$) is comparable with the distribution of
central surface brightness in late type low surface brightness
galaxies ($<\mu_{0}{\rm (LSBG)>_V} \approx 22.7\pm0.7$ mag/${\tt ''}^2$).
These results suggest that the progenitors of HII galaxies may be blue
low surface brightness galaxies and should be detectable in deep
surveys.

The expected number density of these progenitors can be estimated if
we consider a simple star formation history for HII galaxies, as
indicated later by this study, namely that, HII galaxies may have
undergone an initial burst of star formation, followed by an
intermediate episode and they are detected now during the present
event.  Hence,

\[
\phi_{progenitors} = (\frac{\Delta t_{burst}}{t_{age}} n)^{-1}\times 
\phi_{{\rm HII} galaxies}
\]

\noindent
where $\phi_{progenitors}$ is the space density of progenitor galaxies
(in Mpc$^{-3}$), $\Delta t_{burst}$ is the duration of one episode of
star formation, $t_{age}$ is the age of the galaxy and $n$ is the
number of bursts.  Making naive assumptions, based on the present
knowledge, we can take $\Delta t_{burst} \approx 5\times10^6$ yrs, and
that HII galaxies have undergone $n = 3$ bursts in $t_{age} \approx
{\rm few} \times 10^9$ yrs.  The space density of emission line galaxies has
been estimated by Salzer (1989) to be $\phi_{{\rm HII} galaxies} = 0.03 $
Mpc$^{-3}$ (integrated over $-13.5 <$ M$_B < -22.5$).  Thus, we predict a
space density of progenitors to be $\phi_{progenitors} \approx 10$
Mpc$^{-3}$.  This number seems to be rather high at face value. The
comparison of an observed space density of late type LSB galaxies with
this prediction would be desirable.

\subsection{The colour - luminosity relation}

\begin{figure*}
\protect\centerline{
\epsfxsize=4.5in\epsffile[ 18 150 590 600]{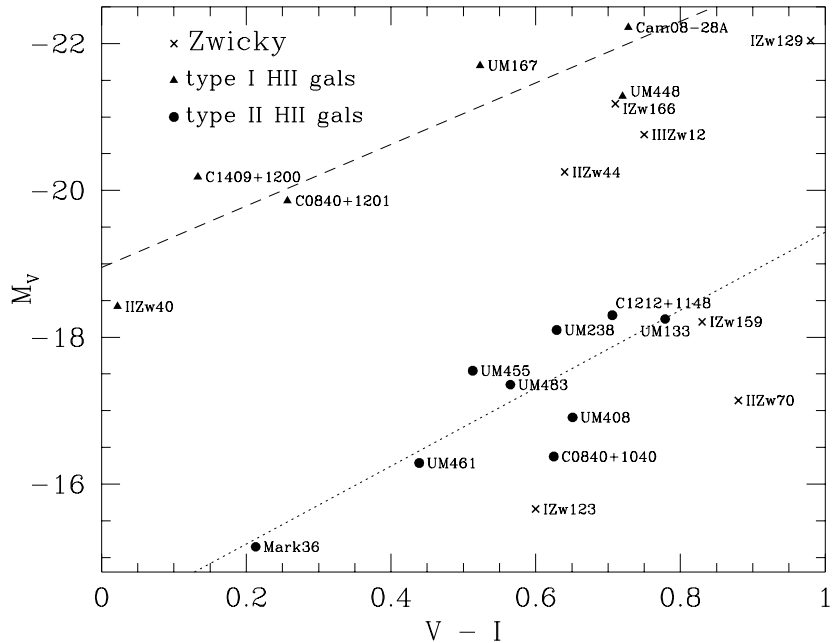}
}
\caption[Total dereddened Colour-Absolute magnitude diagram.]{Total
dereddened Colour-Absolute magnitude diagram. Filled circles are HII
galaxies.  Crosses are Zwicky compact objects.  The lines illustrate
the loci of the apparent two sequences of HII galaxies.  Type I HII
galaxies (with disturbed morphology) fall in the high luminosity
regime (dashed lines). Type II HII galaxies (with symmetric and
regular morphology) fall in the low luminosity regime (dotted lines).}
\label{fig:abs-colour}
\end{figure*}

Figure~\ref{fig:abs-colour} presents the \underline{total} dereddened
V--I colour -- absolute V magnitude diagram of the HII galaxies
(filled circles) in the present sample. The first point in this
diagram is that there is not a single luminosity--colour relation for
HII galaxies when all the points are plotted. However, one can claim
that the brighter objects do not belong to same colour-absolute
magnitude sequence as the objects of luminosity lower than that of
IIZw40 (M$_{\rm V}
\approx -18.5$).  These two apparent sequences are illustrated as
dashed and dotted lines in the figure.

In order to understand the origin of the starburst phenomena in these
low metallicity dwarf systems it is crucial to verify the reality of
these two sequences as well as of the gap between them. This would
allow us to to obtain a homogeneous class of objects and its
observational properties.  This is the whole basis for the
morphological classification schemes of galaxies.  In Telles, Melnick
\& Terlevich (1996) we have proposed a classification scheme based on
the overall morphology of the galaxy; HII galaxies with disturbed
morphology were classified as Type I HII galaxies while Type II HII
galaxies have symmetric and compact morphology.  Typically Type I HII
galaxies are more luminous, have more luminous starbursts and larger
emission line widths.  Furthermore, possibly the most important
characteristics is that the disturbed morphology of Type I systems
(luminous HII galaxies) are suggestive of tidal effects or mergers.
Telles \& Terlevich (1995), however, have shown that these systems are
isolated from bright galaxies. Type II HII galaxies, although they may
have more than one burst, have no apparent signs of being affected by
tidal effects.  

In figure~\ref{fig:abs-colour}, we find that galaxies in the high
luminosity \,``sequence\,'' in this diagram have disturbed morphology
while the ones in the low luminosity \,``sequence\,'' have regular
morphology.  We also show a sub-sample of Zwicky compact objects from
Sargent (1970), with photometry by Moles \etal (1987), which have the
same range of V--I colours as the HII galaxies, that is, the blue
Zwicky compact galaxies (crosses).  The objects of low luminosity
(IZw123, IIZw70, IZw159) are described by Sargent as structure-less,
very compact and no mention to visible companions are made.  The
galaxies of higher luminosity in this range of colours (IIZw44,
IIIZw12, IZw166, IZw129) are described as having faint
\,``jets\,'', plumes, with a companion at some distance.  This seems
to be further support to the classification scheme in two broad
classes of objects.  We believe that this observational hint may
conceal important information on the processes of triggering and star
formation history in dwarf galaxies as a class.

It is therefore tempting to speculate that the mechanisms of
triggering of star formation may be different in the dwarfs from the
higher luminosity systems.  More massive (luminous) bursts may be
formed preferentially from the merging of two gas rich galaxies.  On
the other hand low luminosity bursts may not have been triggered by a
strong direct interaction, but rather induced by {\em internal}
disturbances.  This process may not cause significant morphological
effect on the galaxy since only a small disturbance of the order of
the sound speed would suffice to stir up the ISM to cause cloud-cloud
collisions (Zaritsky \& Lorrimer 1993).  Alternatively, for these Type
II galaxies star formation may have occurred solely due to the secular
dynamical evolution of the primordial collapsing proto-cloud.

\subsection{The connection with other dwarfs}

The similarity of the structural and photometric properties of the
underlying galaxy of HII galaxies with other type of dwarf galaxies
suggests a possible connection among them.  The connection among dwarf
galaxies have long been discussed in the literature.  Different
evolutionary scenarios have been proposed: 1) dE's are remnants of
dI's that have been swept of their ISM (Faber \& Lin 1983). 2) dE's
are quiescent irregulars observed between bursts of star formation
(Gerola, Seiden \& Schulman 1980). 3) Metal poor gas-rich dwarf
irregulars evolve through bursting HII galaxy stage and eventually
fade to become a dwarf elliptical galaxy
(dI$\rightarrow$HII$\rightarrow$dE) (Davies \& Phillips 1988).  Bothun
\etal (1986) and Binggeli (1985) both have shown that bright dI have
too low surface brightness and too long scale lengths to fade into
bright dE's.  Their conclusions argue against (1).  Impey, Bothun \&
Malin (1988) and Bothun \etal (1985) both failed to detect HI gas in
dE's in Virgo.  This result quite convincely rules out (2).

The colours of the underlying galaxies suggest low surface brightness
blue galaxies to be quiescent counterparts of HII galaxies in the
process of accumulating fuel for the intermittent burst of star
formation, until eventual gas depletion.  The structural properties
differ for both ends of the luminosity range in dwarfs.  The same is
valid for HII galaxies which is the basis of our classification
scheme. In Telles (1995) we found that \,``aged\,'' HII galaxies and
dwarf ellipticals fall approximately in the same locus in the
Luminosity - Surface Brightness diagram.  Furthermore, under the
assumptions of mass loss by a \,``normal\,'' IMF population, an
adiabatic dynamical evolution will shift an HII galaxy in the general
direction of the locus of the dwarf ellipticals in the size - velocity
dispersion diagram [R - $\sigma$].  Although the results for the small
sample of HII galaxies do not allow us to discuss in detail, the
trends are suggestive of the existence the of an evolutionary scenario
for dwarf galaxies.

HII galaxies have also been identified as the local counterparts of a
population of Faint Compact Narrow Emission Line Galaxies which may be
associated with the galaxies producing the excess number counts at
redshifts of about 0.3 (Guzm\'an \etal 1996, Koo \etal 1995).  The
nature of these faint blue galaxies is still one of the major
unresolved puzzles and a very active area of research in modern
cosmology (see a review by Koo 1994).  The comparison of the global
properties of these galaxies with those of HII galaxies may help shed
some light on our understanding of their origin and evolutionary
history.

\begin{table*}
\centering
\caption{Models parameters of the interest for comparison.}
\label{tab:model_comp}
\begin{tabular}{cccc} \hline
     & Cervi\~no \& Mas Hesse & Leitherer \& Heckman & Bressan \etal  \\ \hline
Star Formation & Instantaneous burst & Instantaneous burst & Instantaneous burst \\ \hline
IMF & M$_l = 2 \msun$ &  M$_l = 1 \msun$ & M$_l = 0.6 \msun$ \\
    & M$_u = 120 \msun$ & M$_u = 100 \msun$ & M$_u = 120 \msun$ \\
    & $\alpha = 2.35$ & $\alpha = 2.35$ & $\alpha = 2.35$ \\ \hline
Metallicity & 1/20 Z$_\odot$, Z$_\odot$ & 1/10 Z$_\odot$, Z$_\odot$ & 1/20 Z$_\odot$, Z$_\odot$ \\ \hline
Evolutionary & Geneva       & Geneva       & Padova \\
tracks       & group (1993) & group (1990) & group (1994) \\ \hline

\end{tabular}
\end{table*}

\subsection{Comparison with  evolutionary population synthesis models.}

\begin{figure*}
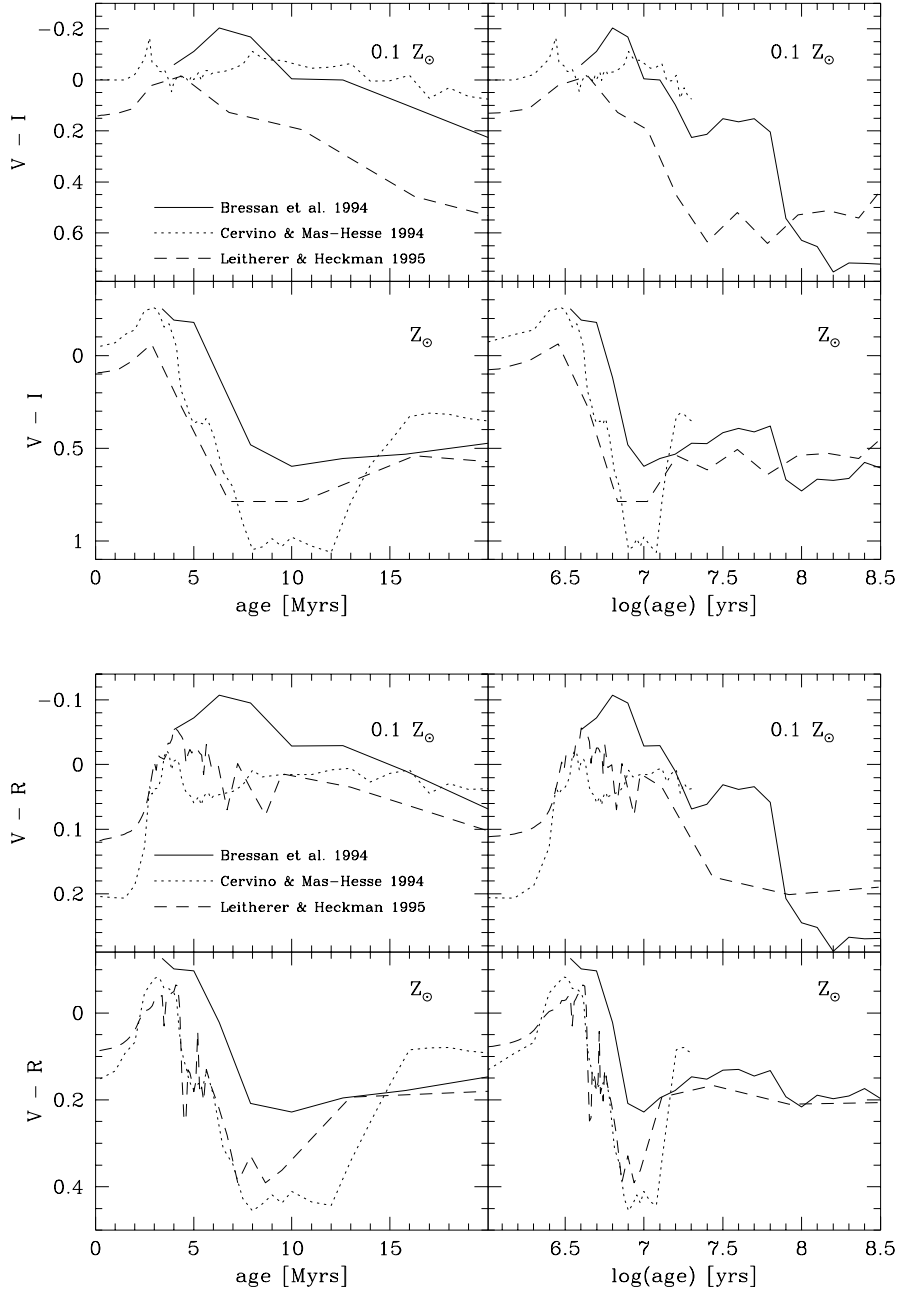

\protect\centerline{
\epsfysize=3.5in\epsffile[ 20 150 580 560]{v_i.models} 
} \protect\centerline{ 
\epsfysize=3.5in\epsffile[ 20 150 580 560]{v_r.models}
} 
\caption[Prediction of recent models for the evolution of a SSP in the optical
V--I and V--R Cousins colours.]{Prediction of recent models for the
evolution of a SSP in the optical V--I (upper panels) and V--R (lower
panels) Cousins colours.  The ages of the models extend to
$2\times10^7$ yrs in the left panels (linear units) and extend to
$3\times10^8$ yrs in the right panels (log units). Model comparisons
are shown for two metallicities (1/10 solar and solar). Different line
types describe models from different authors as shown.}
\label{fig:model_comp}
\end{figure*}

\begin{figure*}
\protect\centerline{
\epsfxsize=6.0in\epsffile[ 30 170 580 710]{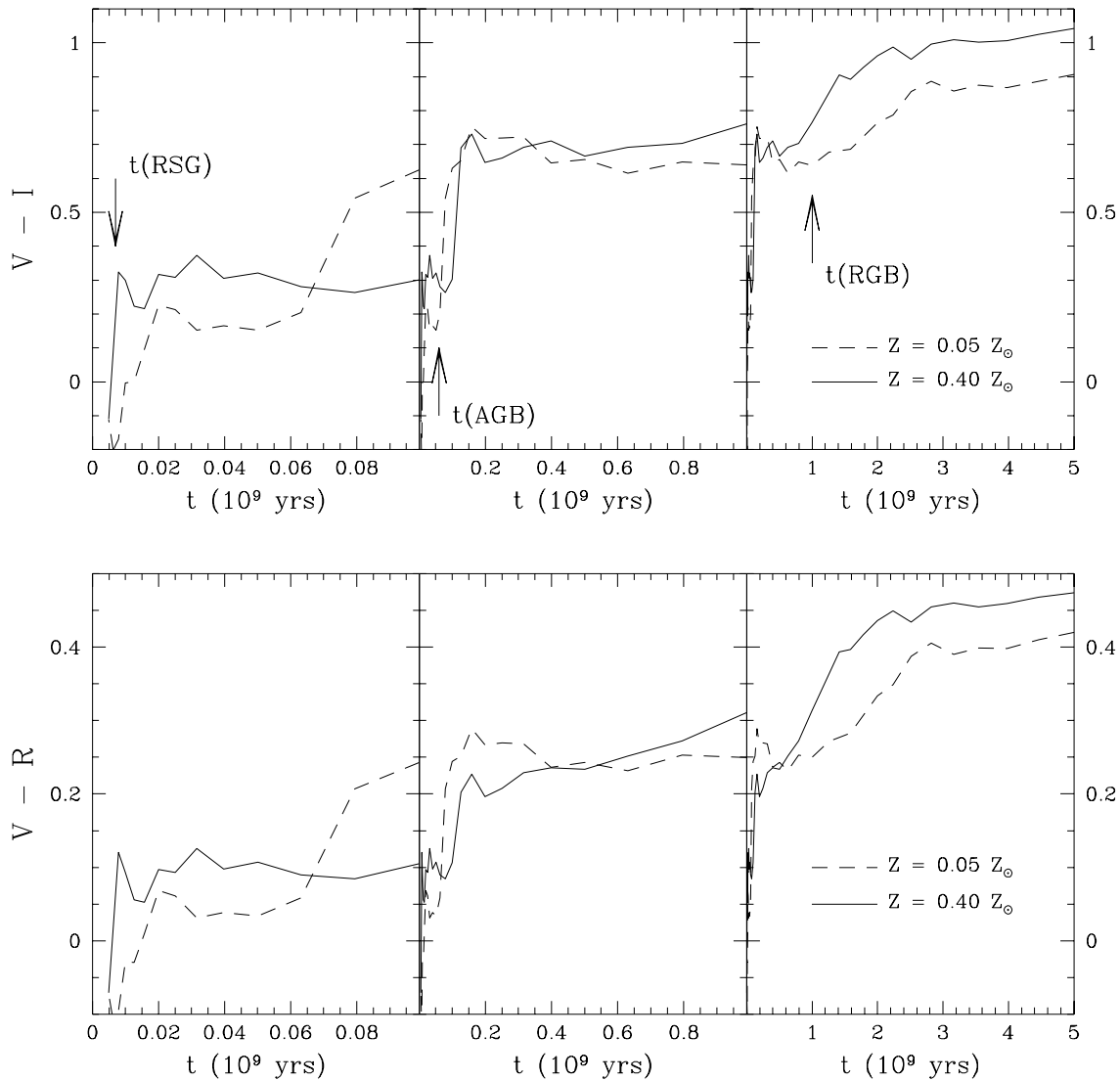}
} 
\caption[ Models of Single Stellar Populations (SSP) from Bressan,
   Chiosi \& Fagotto (1994).]{Models of Single Stellar Populations
   (SSP) from Bressan, Chiosi \& Fagotto (1994). Top panels: V--I
   Cousins colours. Bottom panels: V--R Cousins colours.  The three
   panels for each colours are the same with consecutive blow-up of
   younger age ranges (0-$5\times10^9$, 0-$1\times10^9$,
   0-$1\times10^8$ yrs).  Solid lines are models for $Z=\frac{2}{5}
   Z_\odot$. Dashed lines are models for $Z=\frac{1}{20} Z_\odot$.}
   \label{fig:models}
\end{figure*}

Knowledge of the stellar content is an important prerequisite before
one can answer questions regarding the physical processes relevant for
understanding star formation and the chemical evolution in HII
galaxies.  In order to have a handle on the properties of the stellar
populations, one can compare the observed properties of the galaxies
such as integrated colours, equivalent widths of the spectral lines,
total luminosities, etc, with evolutionary population synthesis models
which are designed to predict such observables.  Slight different
predictions of the observables may arise from different evolutionary
population models from different treatment and/or input parameters
regarding the library of stellar evolutionary tracks used to calculate
isochrones in the theoretical colour magnitude diagrams (CMD), library
of observed stellar spectra to derive the integral spectral energy
distribution which will lead to magnitudes and colours, and the
assumptions for star formation and chemical enrichment.  Therefore,
one needs to choose which model one will adopt to compare with
observations depending on available constraints of interest in each
case.  Very often the results are not unique and only global
properties of the observed galaxy are retrieved.

In this section we have compared the results of three recent
evolutionary models for an episode of star formation of no duration
which synthesize broad band colours of the immediate
interest. Firstly, we have chosen two models (Cervi\~no \& Mas Hesse
1994, Leitherer \& Heckman 1995) which describe in detail the early
evolution of the starburst with a short time resolution ($\Delta \sim
0.1$ Myrs).  These models are based on the most up-to-date input
physics for the theory of stellar atmospheres, stellar winds, and
stellar evolution.  Both models also include nebular continuum
emission but do not consider gaseous line emission.  They are,
however, optimized for the treatment of the massive stars in the
starburst.  As discussed by these authors, massive stars ($M > 5~
M_\odot$ in the Zero Age Main Sequence) evolve in only a few million
years, therefore it is reasonable to assume that they have all formed
coevally in an instantaneous burst of star formation.  Other
observational properties also put tight constraints on the age and
composition of the stellar populations in HII galaxies.  The
equivalent width of H$\beta$ has been shown to be an age indicator for
young bursts of star formation (Copetti, Pastoriza \& Dottori 1986).
The typically large values for this observable in HII galaxies
indicate that the present burst of star formation must be younger than
a few 10$^7$ years.  In addition, in galaxies where the Wolf-Rayet
(WR) star signatures are detected, the inferred large ratio of the
number of Wolf-Rayet stars to the total number of massive stars that
undergo a Wolf-Rayet phase [WR/(WR+O)] provide a strong evidence
towards the relatively short duration of these intense star formation
processes, since they can not be reproduced assuming a constant star
formation rate (Vacca \& Conti 1992).  For these reasons we only
consider Single Stellar Population (SSP) models in our analysis.

For the prediction of the evolution of the stellar system after all
ionizing stars have exploded as supernovae we have made use of the
detailed chemical-spectro-photometric models of population synthesis
by Bressan, Chiosi \& Fagotto (1994) who include all evolutionary
phases, from the main sequence till late stages of stellar evolution.
Bressan, Chiosi \& Fagotto models (1994) do not include the nebular
continuum emission, which in any case is not significant in the
optical regime, nor gaseous line emission.  However, since it was
designed to represent the evolution of the stellar population of
elliptical galaxies it follows the evolution to older ages.  This will
be useful when we want to search for the underlying older stellar
populations by comparing with the colours of the extensions of the
sample of HII galaxies.

Table~\ref{tab:model_comp} shows some of the basic parameters for the
models in question.  Figure~\ref{fig:model_comp} shows the predictions
of the different models for the optical (V--I and V--R) Cousins
colours of a starburst for two metallicities ($\sim 1/10$ solar,
solar). In all cases the published Johnson colours have been
transformed to the observed Cousins photometric system using the
formulae given in Bessell (1979).  Agreement is reasonably good
between Leitherer \& Heckman's models (dashed lines) with Cervi\~no \&
Mas Hesse's (dotted lines) who use the old and the new, respectively,
evolutionary tracks from the Geneva group.  They are very similar in
V--R at both metallicities, but disagree to up to $\sim 0.4$ mag in
V--I at low metallicities.  Bressan, Chiosi \& Fagotto models which
are not optimized to describe the evolution of the starburst, although
it uses all evolutionary phases according to the Padova tracks, are
only in general agreement with the other models. The single largest
discrepancy seems to be associated with the lack of a strong
supergiant burst in the Padova colours at a few million years, notably
at solar metallicity.

In order to \,``disentangle\,'' ages and metallicity of the
synthesized stellar population models we have assumed the gas
abundances derived from the emission line fluxes to be a reasonable
upper limit of the metallicity of the underlying population.
Therefore, we are only concerned with the models of low metallicity (Z
$\sim$ 1/10 Z$_\odot$).  For this regime, the upper panels for both
colours of figure~\ref{fig:model_comp} show that these optical colours
alone are not very sensitive to the early evolution of the starburst
($t < 2\times10^7$ yrs).  This trend is followed by any of the three
models.  One can only say that a low metallicity starburst with
colours bluer than V--R $< 0.1$ {\em and} V--I $< 0.3$ will be
dominated by a single stellar population younger than a few $10^7$
yrs.


Figure~\ref{fig:models} shows the colour evolution of a SSP for two
different sub-solar compositions (Z = 0.001 as dashed lines and Z =
0.008 as solid lines; in this unit Z$_\odot$ = 0.02) from Bressan,
Chiosi \& Fagotto (1994). Upper panels show the predictions of V--I and
lower panels the predictions for V--R Cousins colours.  The results
refer to the Salpeter IMF with $m_l = 0.85$ \msun~and $m_u = 120$
\msun.  Details of the construction of the models can be found in
their paper (Bressan, Chiosi \& Fagotto 1994).  The panels from the
left to the right show different age ranges of the same models in
order to illustrate the three main phases which characterize the
evolution of a SSP.  These may be called \,``phase transitions\,''
which cause the models to predict sudden changes to the colours as the
evolution proceeds at the appearance of particular evolutionary phases
of different types of star.  These \,``jumps\,'' to redder colours are
most remarkable at the appearance of the asymptotic giant branch (AGB)
at about $10^8$ yrs and the red giant branch (RGB) stars at about
$10^9$ yrs (see Table 1 in Bressan, Chiosi \& Fagotto).  The AGB and
RGB phases can be experienced by stars of different initial masses and
lifetimes, hence characteristic ages.  These ages are shown as arrows
in figure ~\ref{fig:models}.

The small difference in chemical composition of the models shown in
figure~\ref{fig:models} at the low metallicity regime does not change
the predictions of these optical colours by very much.  For $Z=1/20
Z_\odot$ (dashed lines) the colours evolve in a slightly bluer path
than at $Z=2/5 Z_\odot$ (solid lines).  The main difference occurs at
very early stages of the evolution (at $\sim 10^7$ yrs) with the
appearance of the first red supergiant phase (RSG) of massive stars.
At low metallicities, helium ignition occurs at hot effective
temperatures for stars with higher initial masses.  These stars start
burning helium as blue supergiants (BSG) and become RSG stars only at
the end of the helium burning phase (Garcia-Vargas, Bressan \&
D\'{\i}az 1995).  For these reasons the model for Z=0.001 (1/20
Z$\odot$), dashed lines in figure~\ref{fig:models}, do not produce the
small peak due to the RSG phase at about $6\times10^6$ yrs as do the
models for higher metallicity.

\begin{figure*} 
\protect\centerline{ 
\epsfxsize=6.5in\epsffile[ 20 150 585 710]{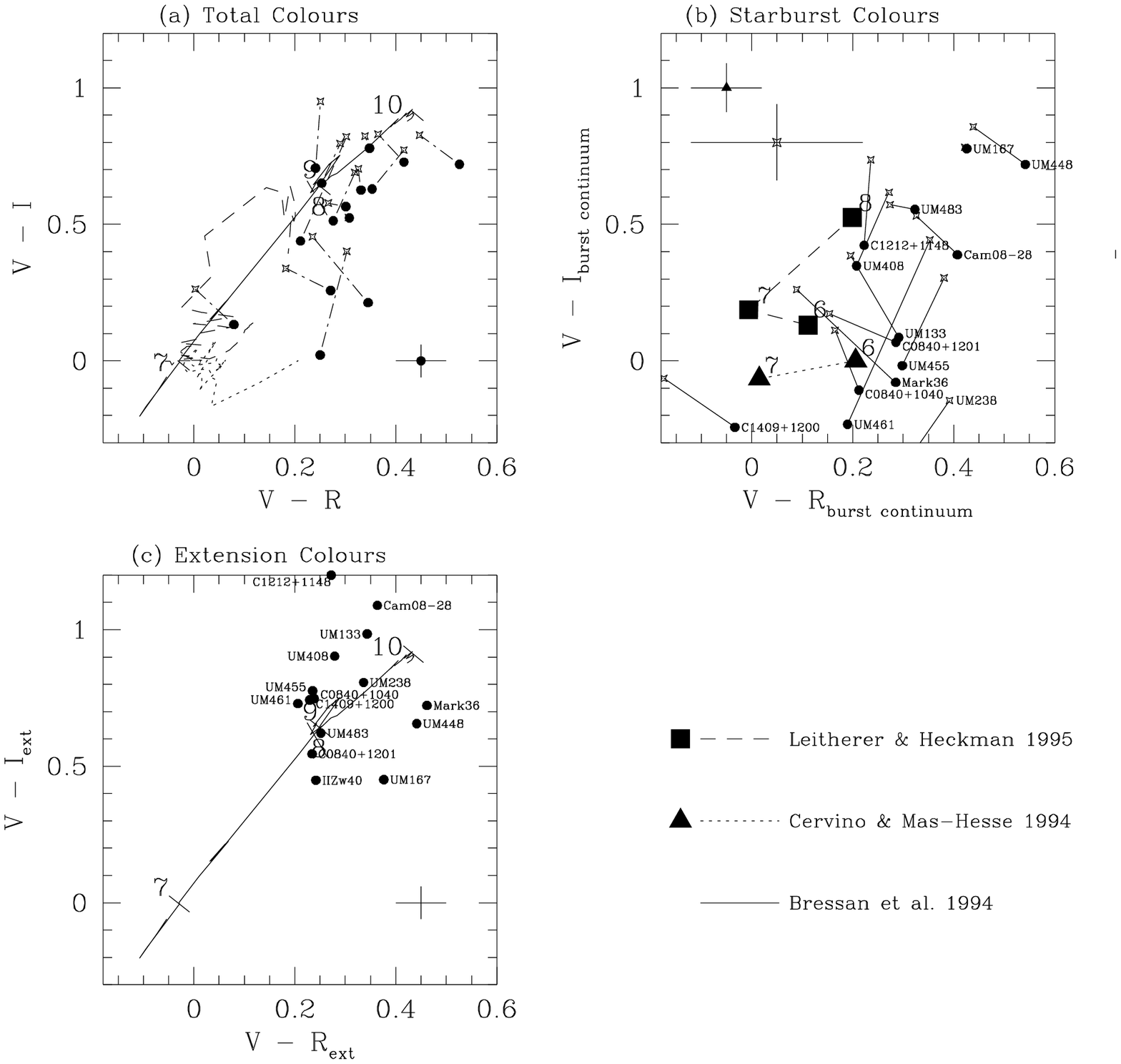} 
} 
\caption[Dereddened colour-colour diagrams.]{Dereddened
   colour-colour diagrams. a) {\em Total} observed colours.  b)
   Colours of the starburst region alone.  In these panels we draw
   lines joining the uncorrected starburst colours (filled circles) to
   the corrected colours (crosses) to illustrate the effect on these
   colours when we subtract the emission line contribution.  c)
   Colours of the extensions.  The stellar evolutionary models are
   shown in the figure.  Labels (numbers) are {\em log}(age).}

\label{fig:colour-colour} 
\end{figure*}

\subsection{Comparison of HII galaxy colours with  evolutionary
population models}

To set limits on the star formation histories in these star forming
dwarfs, we have compared Bressan, Chiosi \& Fagotto 's models with the
colours of the present sample of HII galaxies.

Figure~\ref{fig:colour-colour} a.b.c show the de-reddened
colour-colour diagrams for the total colours (integrated colours of
the whole galaxy), for the starburst colours and for the colours of
the extensions, respectively (see
Section~\ref{stellar:burst_galaxy_col} for a description of how these
two components are separated).   The colours of the extensions are
corrected for galactic extinction only.  The colours of the starburst
are corrected for total line of sight extinction using the estimate
from the Balmer decrement [C(H$\beta$)] derived from the
spectrophotometric information in SCHG, assuming case B recombination.
The typical estimated internal photometric errors are shown in each
panel as a solid line cross.  The dashed line error cross in
figure~\ref{fig:colour-colour}b corresponds to the estimated error in
the emission line subtracted colours of the starburst (starburst
continuum colours).

Figure~\ref{fig:colour-colour}a shows the colour-colour diagram for
the integrated colours of the whole galaxy corrected for galactic
extinction. The dot-dashed lines in this figure show the direction to
which the correction for the contribution of emission lines in the
broad band filters will shift the total colours.  The crosses
represent the colours after this correction has been applied. The
correction for line emission applied here is described in Appendix A.
In this panel the solid line represents the single stellar population
(SSP) models at 1/20th of solar metallicity (Z=0.001) from Table 3 of
Bressan, Chiosi \& Fagotto (1994).  The dotted line is the starburst
model of Cervi\~no \& Mas-Hesse (1994) and the dashed line is the
starburst model of Leitherer \& Heckman (1995) as shown.  It is clear
from this top left panel that there is an inconsistency between the
stellar population models and the observed colours of HII galaxies.
The inconsistency still remains even after subtraction of the
contribution of the emission lines within the starburst region.
Therefore, we have separated the starburst from the extensions of each
galaxy in order to have a better handle on the colours of the two
distinct stellar population components.  Thus, we have attempted to
put constraints on the age of the stellar population both of the
stellar system within the starburst and the underlying stellar system
by comparing these derived colours with an adopted evolutionary model.
From what follows, we have adopted models with metallicity as near as
possible to the upper limit imposed by the observed heavy element
abundance derived from the nebula emission.  Therefore, the models are
all low metallicity (Z = 0.1 \Zsun).
 
\subsubsection{The age of the starburst} 
 
Figure~\ref{fig:colour-colour}b shows the observed colours of the
starburst (filled circles).  The dashed lines again, represent the
shift of the points in this diagram caused by the subtraction of the
flux contribution in the broad band filters by the nebular emission
lines ([OIII], H$\beta$ and H$\alpha$).  Thus, the tip of the line
represents the starburst continuum colours.  We have also subtracted
the contribution of the underlying galaxy to the colours within the
starburst region, assuming a constant surface brightness of the
underlying galaxy as given in Table~\ref{tab:extcolours}.

From this figure one can see that most of the starbursts have colours
that fall significantly far from the locus of the colours predicted
for very young single stellar bursts ($\approx 10^7$ yrs), considering
the constraints on the age and duration of the burst by the
spectroscopic observations (Coppetti, Pastoriza \& Dottori 1986; Vacca
\& Conti 1992; Cervi\~no \& Mas-Hesse 1994; Garcia-Vargas, Bressan \&
D\'{\i}az 1995). Shown in this figure as filled squares is the model
by Leitherer \& Heckman (1995) from 10$^6$ to 10$^8$ yrs and as
filled triangles the models by Cervi\~no \& Mas-Hesse (1994) from
10$^6$ to 10$^7$ yrs.  As can be seen from table~\ref{tab: colours},
the starburst colours of IIZw40 fall outside the range plotted here.
This is a very low galactic latitude object and the extinction
correction is relatively more uncertain.

At this point we are unable to derive definitive conclusions from the
broadband optical colours alone about the stellar content in the
starburst in HII galaxies.  Various sources of uncertainties may be
the causes of the discrepancies.  Some of them are:

\newcounter{bean1}

\begin{list}
{(\roman{bean1})}{\usecounter{bean1}}

\item Reddening: We may be strongly underestimating the reddening of
the stellar population within the starburst by using the estimate from
the emission line ratio and assuming case B recombination.

\item Underlying galaxy: The assumption of a \,``sheet-like\,''
underlying galaxy of constant surface brightness may be a poor one.
We may be underestimating the contribution of a underlying population
of stars formed previous to the present burst.

\item The emission line correction: An assumption of a constant surface
brightness distribution of the emission lines is rather crude.  A
better account for the emission line contribution and its surface
brightness distribution can be achieved in an analogous way to the
procedure here with accurate narrow band surface photometry centered
on either H$\alpha$ or [OIII] lines.

\item Models: The late stages of massive star evolution may not be well
prescribed by the models, mainly at low metallicities.
The uncertainties include the photoionization and energy output from
massive stars, the relative numbers of red and blue supergiants, the
evolution of Wolf-Rayet stars and the rate of Type II supernovae, as
emphasized by recent comparative studies of evolutionary synthesis
models by Charlot, Worthey \& Bressan (1996) and Garc\'{\i}a-Vargas,
Bressan \& Leitherer (1996).
\end{list}

The role of many parameters in stellar evolutionary population
synthesis models are still worthy of close scrutiny and detailed
study.  But from the observational point of view, we believe there are
still methodologies (like the one presented here) of data analysis
that can be improved and devised in order to probe the stars in
starburst.

\subsubsection{The age of the underlying galaxy} 
 
Figure~\ref{fig:colour-colour}c shows the colour-colour diagram for
the extensions of HII galaxies.  These regions are beyond the ionized
region and represent the underlying population formed in previous
episodes of star formation. The position of the galaxy colours in this
diagram seems to suggest that the underlying galaxy in HII galaxies
have ages ranging from a few 10$^9$ to 10$^{10}$ years when compared
with the SSP models with standard IMF at low
metallicity of Bressan, Chiosi \& Fagotto (1994). These are
indications, therefore, of an intermediate age population ($\sim 10^9$
yrs) in HII galaxies.  

This brief stellar population analysis is consistent with the findings
of the comparison with the colours of other dwarf galaxies and seems
to rule out the young galaxy hypothesis for these systems. They are
unlikely to be experiencing their very first burst of star formation.
 
\subsubsection{A composite model for the underlying galaxy} 
 
\begin{figure} 
\protect\centerline{ 
\epsfxsize=3in\epsffile[ 30 170 580 710]{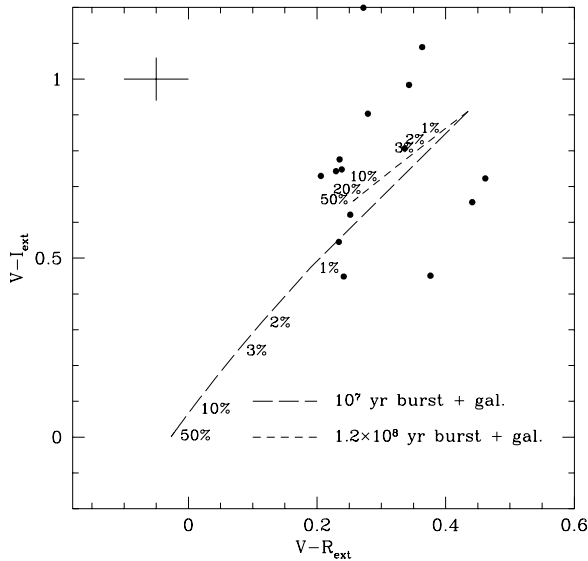} 
} 
   \caption{Composite models. Extension colours} 
   \label{fig:composite} 
 \end{figure} 
 
We have investigated the properties of the underlying galaxy deduced
from the colours beyond the region which is presently undergoing
violent star formation (the extensions) in order to infer what one
expects to observe in the quiescent periods of HII galaxies.
Figure~\ref{fig:composite} shows two SSP composite models, as a
function of the fraction of the total mass involved in the burst (the
strength of the burst), one representing a burst of $10^7$ yrs old
(long dashes) superposed on a $10^{10}$ yr old underlying galaxy, the
other a burst of $1.2\times10^8$ yrs old (short dashes) superposed on
a $10^{10}$ yr old underlying galaxy. A recent episode of star
formation ($10^7$ yrs old) superposed on an old galaxy would make the
colours of the underlying galaxy much bluer than observed even if only
a small fraction of the total mass participated in this event (long
dashes).  On the other hand, if no other burst occured since the very
first episode $10^{10}$ yrs ago the colours would have remained redder
than observed.  The fact that the observed colour of the extensions
lie roughly between the locus of the evolution of a composite system
with a $1.2 \times 10^8$ yr old event and a pure old galaxy (short
dashes) suggests that HII galaxies may have undergone an intermediate
episode of star formation approximately $10^9$ yrs ago and the light
now is dominated by the intermediate mass stars in the AGB phase.
Although we do not claim that these results are conclusive, they seem
to be compatible with the view of episodic star formation followed by
long quiescent periods for HII galaxies.
 
Separating the burst and extension has improved the compatibility with
the models for the extension.  However, the models remained
inconsistent with the observed colours of the burst.  The method
applied here, however, may be further improved and useful specially
with deeper observations and a wider spectral dynamical range.
Ferguson (1994) has also compared the B-V and V--I colours of a sample
of different types of dwarf galaxies from the literature with the same
evolutionary models used here (Bressan, Chiosi \& Fagotto 1994) as
well as with models of Worthey (1994).  He has shown that the observed
colours are incompatible with the expectations from these stellar
population models.  Therefore, until further tests of the calibration
of the galaxy colours and the uncertainties and application of the
models are better understood, we will simply take these results as
tentative.  These results show that if models do not fit the
observations of present day \,``young\,'' systems, there seems to be
little hope for studies of young systems at high $z$ where additional
uncertainties are present (e.g. K corrections, lack of spatial
resolution or spectrophotometric information).

\section{Conclusions}\label{stellar:conclusions}

We have presented evidence from a morphological, photometric and
structural point of view that HII galaxies show a wide range of
morphologies with their own characteristics, despite of sharing the
common property of having a dominant giant HII region. The very cause
of the enhanced star formation or possible mechanisms which may have
triggered the starburst may have been different in different types of
objects.  While in more compact single systems the internal dynamics,
stochastic star formation or propagating star formation may have a
major role, for more luminous systems mergers may have had their act
in play.

Some of the main conclusions are:
\begin{list} 
{(\roman{bean1})}{\usecounter{bean1}}

\item We have confirmed the adequacy of the broad classification
scheme, devised in Telles, Melnick \& Terlevich
(1996), namely that, HII galaxies may be described as two different
classes of objects:

\begin{itemize}
\item Type I HII galaxies ($M_{\rm V} < -18.5$) are luminous, have
larger emission line widths and have marginally higher heavy metal
abundances, and show signs of being interacting or merging systems.
The most luminous Type I's also have small equivalent widths of
W(H$\beta$).
\item Type II HII galaxies ($M_{\rm V} > -18.5$) are compact and
regular. They tend to be marginally more metal deficient and show no
signs of being products of interactions or mergers.
\end{itemize}

\item The optical colours of the underlying galaxies are similar to the
colours of late type low surface brightness galaxies. These may be
good candidates for being the quiescent counterparts of HII galaxies.

\item Evolutionary models agree only qualitatively among themselves.  
They, however, fail to fit quantitatively the observed colours of the
starburst in low metallicity HII galaxies to a few tenths of
magnitude.  We fear that the predictions of the models for the
evolution of young galaxies at high redshift may result in even larger
uncertainties.

\item The comparison of the observed colours with models of
evolutionary population synthesis models are inconclusive.  This is a
consequence of uncertainties both in the observed colours and perhaps
more seriously, in the models themselves.  In any case, if the models
are right for old stellar systems at low metallicity, the colours of
the underlying galaxy in HII galaxies are not compatible with them
being truly young galaxies having their first burst of star formation.
They have indicated the possible presence of an intermediate age
population of approximately 1 Gyr ago, suggestive of an episodic mode
of star formation for HII galaxies.  Surface photometry in the
near-infrared, together with the results presented here, may help put
more strict constraints on the star formation history in HII galaxies.


\end{list}

\section*{Acknowledgments}  
  
ET acknowledges his grant from CNPq/Brazil.  We thank Harry Ferguson,
Stacy McGaugh \& Bernard Pagel  for a critical reading of the original
version of this paper and for valuable discussion on this work.  
We also thank Miguel Cervi\~no, Claus Leitherer and Alessandro Bressan for
providing the  results of their models in computer form.

\appendix

\section{Correction  for emission-line contamination}\label{app:line_cor}

The approach used here for the correction for the contamination by
emission lines of the broad band CCD images follows closely the one
used by Salzer, MacAlpine \& Boroson (1989) with only a difference on
the size factor and the fact that here we deal with the V, R and I
bandpass and exclude B as described below.

The colours of a galaxy with strong gaseous emission  may be significantly
different from the continuum-only colour.  For objects with small redshift,
H$\beta$ and [OIII]$\lambda\lambda$4959,5007 will be redshifted into the V
bandpass and H$\alpha$ will fall well into the R bandpass.  The I bandpass,
however, is virtually unaffected due to the absence of comparatively
strong gaseous emission and also due to its broader filter width.

The contribution of the emission lines to the total flux is then corrected
in the following manner:

\[
{\cal I}_{Total}(\lambda,r) = {\cal I}_{gas}(\lambda,r) + {\cal I}_{continuum}(\lambda,r)
\]
\noindent
where ${\cal I}$ is the intensity at a given wavelength and distance from the center. Hence,

\[
{\cal F}_{Total}^i(r) = \int_{i} {\cal I}_{Total}(\lambda,r) \times T_i(\lambda) d\lambda \]
\[
= \int_{i} {\cal I}_{g}(\lambda,r) \times T_i(\lambda) d\lambda + \int_{i} {\cal I}_{c}(\lambda,r) \times T_i(\lambda) d\lambda
\]
\[
{\cal F}_{Total}^i(r) = {\cal F}_{g}^i(r) + {\cal F}_{c}^i(r)
\]
\noindent
where ${\cal F}_{Total}^i(r)$ is the total flux in filter $i$ and $T_i(\lambda)$ is the combined normalized CCD sensitivity and filter transmittance at a
wavelength $\lambda$.

For the contribution of an emission line $l$, we can describe the intensity by

\[
{\cal I}_{g}(\lambda,r) = {\cal F}_{g}^i(r) \times \delta(\lambda-\lambda_l)
= {\cal I}_{c}(\lambda,r) \times W_l(r) \times \delta(\lambda-\lambda_l)
\]
\noindent
$W_l(r)$ is the equivalent width of line $l$.
Hence,

\[
{\cal F}_{g}^i(r) = \int_{i} {\cal I}_{c}(\lambda,r) \times W_l(r)
\times \delta(\lambda-\lambda_l)\times T_i(\lambda) d\lambda \]
\[
 = {\cal I}_{c}(\lambda_l,r) \times W_l(r) \times T_i(\lambda_l)
\]
and now $T_i(\lambda_l)$ is the value for the combined normalized
 CCD sensitivity and filter transmittance at
{\em redshifted} wavelength of the emission line $l$.

Assuming a plane continuum [${\cal I}_{c}(\lambda,r) =
{\cal I}_{c}(r)$] one would have for its contribution to
the total flux the following:

\[
{\cal F}_{c}^i(r) =  {\cal I}_{c}(r) \times \Delta\lambda_i
\]
\noindent
where $\Delta\lambda_i = \int_{i} T_i(\lambda) d\lambda$ is the filter bandpass
width.
Finally,

\[
{\cal F}_{Total}^i(r) = {\cal I}_{c}(r) \times
W_l(r) \times T_i(\lambda_l) + {\cal I}_{c}(r) \times \Delta\lambda_i \]
\[
= {\cal F}_{c}^i(r) \times {\sum [\frac{W_l(r)}{\Delta\lambda_i} \times T_i(\lambda_i)] + 1}
\]

\noindent
where the $\sum$ indicates that one such term for each line $l$
considered in filter $i$ should be included.

Salzer, MacAlpine \& Boroson (1989) introduce a multiplicative factor
${\cal R}$ in the sum which is the ratio of the area of the slit to
the area where their broad and measurements were made,

\[
{\cal F}_{c}^i = \frac{{\cal F}_{Total}^i}{[\sum \frac{W_l}{\Delta\lambda_i} \times T_i(\lambda_i) \times {\cal R}] + 1}
\]

As opposed to Salzer, MacAlpine \& Boroson (1989) though we are only
correcting the flux within the burst region of a galaxy, not the total
galaxy flux, which has been separated by subtracting the narrow band
CCD image from the broad band ones to discriminate the region of
gaseous emission, thus for us $\cal R =$ 1.  This does not solve the
problem of the spatial variation of surface brightness including the
variation of the strength of the lines W(H$\beta$) across the burst
region.  It is worth noting as well that the correction for the V--R
colours, just as the B-V colours of Salzer, MacAlpine \& Boroson are
small because they are differential between {\em both} fluxes.
However, the correction in V--I and R--I colours will be much larger
since there is no correction in the I fluxes.  Correction for the
total colours of the galaxy will be larger the larger the contribution
of the burst region to the total fluxes, or similarly the larger the
ratio of burst area to total area.

\end{document}